\documentclass[aps,prc, reprint,showpacs,floatfix]{revtex4-1}
  
\usepackage[colorinlistoftodos,prependcaption]{todonotes}
\usepackage{graphicx}
\usepackage{epstopdf}
\usepackage{amssymb}
\usepackage{booktabs}
\usepackage{amsmath}
\usepackage{textcomp}
 
\usepackage{xcolor} 
\usepackage{color}
\usepackage{tabularx}
\usepackage{import}

\newcommand{\eightbe}{$^{8}$Be}
\newcommand{\ninebe}{$^{9}$Be}
\newcommand{\bi}{$^{209}$Bi} 
\newcommand{\pb}{$^{208}$Pb} 
\newcommand{\pt}{$^{196}$Pt} 
\newcommand{\w}{$^{186}$W}
\newcommand{\er}{$^{168}$Er} 
\newcommand{\sm}{$^{144}$Sm} 
\newcommand{\zeroplus}{$0^{+}$}
\newcommand{\twoplus}{$2^{+}$}
\newcommand{\platypus}{{PLATYPUS}}
\newcommand{\balin}{BALiN}
\newcommand{\erel}{$E_{\mathrm{rel}}$}
\newcommand{\qval}{Q-value}
\newcommand{\thetaonetwo}{$\theta_{12}$}
\newcommand{\heavy}{\sm, \er, \w, \pt, \pb\ and \bi}  
 
\newcommand{\thetabeta}{$\theta_{12}-\beta$}
\newcommand{\fivehalf}{$\frac{5}{2}^-$}
\newcommand{\sigmacf}{$\sigma_\mathrm{CF}$}
\newcommand{\sigmaicf}{$\sigma_\mathrm{ICF}$}
\begin{document}


\title{Importance of lifetime effects in breakup and suppression of complete fusion in reactions of 
weakly bound nuclei}

\author{K.J. Cook}
\email{kaitlin.cook@anu.edu.au}
\author{E.C. Simpson}
\author{D.H. Luong}
\author{Sunil Kalkal}
\author{M. Dasgupta}
\author{D.J. Hinde}
\affiliation{Department of Nuclear Physics, Research School of Physics and
Engineering, The Australian National University, Acton ACT 2601, Australia}

\date{\today}

\begin{abstract}
\begin{description}
 \item[Background]
 Complete fusion cross sections in collisions of light, weakly bound nuclei and high Z targets 
show suppression of complete fusion at above-barrier energies. 
This has been interpreted as resulting from breakup of the weakly bound 
nucleus prior to reaching the fusion barrier, reducing the probability of complete charge capture. 
Below-barrier studies of reactions of $^9$Be have found that breakup of \eightbe\ formed by 
neutron stripping dominates over direct breakup, and that transfer triggered breakup may account 
for the observed suppression of complete fusion. 
 \item[Purpose] This paper investigates how the above conclusions are affected by lifetimes 
of the resonant states that are populated prior to breakup. If the mean life of a populated 
resonance (above the breakup threshold) is much longer than the fusion timescale, then its 
breakup (decay) cannot suppress complete fusion. For short-lived 
resonances, the situation is more complex. This work explicitly includes the mean life of the 
short-lived \twoplus\ resonance in \eightbe\ in classical dynamical model calculations to 
determine its effect on energy and angular correlations of the breakup fragments and on model 
predictions of suppression of cross sections for complete fusion at above-barrier energies.
\item[Method] Previously performed coincidence measurements of breakup fragments 
produced in reactions of \ninebe\ with \heavy\ at energies below the barrier 
 have been re-analysed using an improved efficiency determination of the \balin\ detector array.
Predictions of breakup observables and of complete and incomplete fusion at 
energies above the fusion barrier are then made using the classical dynamical 
simulation code \platypus, modified to include the effect of lifetimes of resonant states.
 \item[Results] The agreement of the breakup observables is much improved when lifetime effects are 
included explicitly. Sensitivity to sub-zeptosecond lifetime is observed. The predicted suppression 
of complete fusion due to breakup is nearly independent of $Z$, and has an average value of $\sim 
9 \% $. This is below the experimentally determined fusion suppression which is typically $\sim 
30\%$ in these systems.
 \item[Conclusions]
Inclusion of resonance lifetimes is essential to correctly reproduce breakup observables.  
This results in a larger fraction of nuclei remaining intact at the fusion barrier radius, compared 
with calculations that do not explicitly include lifetime 
effects. The more realistic treatment of breakup followed in this work leads to the conclusion that 
the suppression of complete fusion cannot be fully explained by breakup prior to reaching the 
fusion barrier. Only one third of the observed fusion suppression can be attributed to the 
competing process of breakup. Other mechanisms that can suppress complete fusion 
must therefore be investigated. One of the possible candidates in cluster transfer that produces 
the same heavy target-like nuclei as those formed by incomplete fusion.   
\end{description}
\end{abstract}


\maketitle

\section{Introduction}
\label{Introduction}
The causes of complete
fusion suppression in above-barrier reactions with light, weakly bound nuclei is a key question 
in fusion dynamics. Fusion measurements of 
\ninebe\ +\pb,\bi \cite{dasgupta1999fusion,Dasgupta2004,Signorini2004,Dasgupta2010} 
and $^{6,7}$Li + \bi\ \cite{Dasgupta2004,Dasgupta2002} show that above-barrier complete fusion 
cross sections (experimentally defined as capture of the full charge of the projectile) are reduced 
by $\sim 30\%$, both in comparison with those predicted by 
complete fusion models and with measurements for well-bound nuclei forming the same 
compound nucleus \cite{Dasgupta2004,Rath2009}. Complete fusion suppression in reactions with 
\ninebe\ has been observed 
for a variety of targets in the range $39\leq Z \leq 83$ 
\cite{Gomes2006,Dasgupta2010,Palshetkar2010,Parkar2010,Fang2013}. This suppression was initially 
suggested to result from direct breakup of 
\ninebe($\rightarrow \alpha + \alpha +n$) prior to reaching the fusion barrier
\cite{dasgupta1999fusion}. It was conjectured that breakup reduces the probability of the full 
charge of the projectile-like nucleus being captured, thus suppressing complete fusion (CF), and 
increasing the incomplete fusion (ICF) cross-sections. 

Experiments were undertaken to probe the 
extent of the role of breakup in complete fusion suppression. These experiments were performed at 
below-barrier energies to allow clearer investigation of breakup, as there is essentially no 
absorption of the charged fragments \cite{Hinde2002}. These investigations found that transfer 
followed by breakup contributes much more than direct breakup to the total breakup probability 
\cite{Rafiei2010b,Luong2011b}. In the case
of \ninebe, breakup in interactions with \heavy\ is dominated by neutron stripping forming 
\eightbe\ which subsequently breaks up into $\alpha + \alpha$, rather than \ninebe\ undergoing 
direct 
breakup into $\alpha + \alpha + n$ or $^8$Be$+n$. 

It was recognised early on \cite{Hinde2002} that very long-lived 
states, such as the \zeroplus\ ground-state of \eightbe, which has a mean life of $\sim 10^{-16}$ s 
\cite{Tilley2004}, results in breakup far from the target-like nucleus. It therefore cannot 
contribute to complete fusion suppression. At above-barrier 
energies, the \eightbe\ nucleus in its 
ground-state will pass inside the fusion barrier and be absorbed before decay 
can occur. However, population of broad resonances with much shorter 
mean lives will result in breakup close to the target-like nucleus.

The question then is: what is the quantitative contribution of near-target transfer-triggered 
breakup to the suppression of complete fusion? This was previously addressed by first obtaining 
breakup probabilities as a function of distance of closest approach (``breakup 
functions'') \cite{Rafiei2010b} at below-barrier energies. These breakup 
functions were then used as input to the 
classical dynamical model code \platypus\ \cite{Diaz-Torres2007,Diaz-Torres2011}, to predict 
complete and incomplete fusion cross-sections at
above-barrier energies \cite{Diaz-Torres2007, 
Rafiei2010b} that agreed satisfactorily with experimental results 
\cite{Gomes2006,Dasgupta2010,Dasgupta2004,Fang2013}.


In \platypus, the lifetimes of the states populated were not explicitly taken into account. 
However, locations of breakup and the lifetimes of states are intimately 
related: finite but small mean lives will change the positions at which breakup occurs 
along the trajectory of the nuclei. Indeed, recent work 
\cite{simpson2016} has highlighted that the precise location of breakup relative to the 
target-like nucleus is critical to reaction outcomes, and further, that there exist experimental 
observables that can probe these effects. 

In this work, we investigate quantitatively the effect of the 
lifetime of short-lived resonant states on breakup processes and the resultant incomplete fusion. 
Measurements of transfer reactions populating \eightbe\ can be completely explained by the 
population of \eightbe\ in its \zeroplus, \twoplus, and at higher excitations, $4^+$ states 
\cite{Barker1962,Barker1988}. In breakup 
following $^7$Li collisions with $^{58}$Ni, it has been shown that transfer populates the 
\zeroplus\ and \twoplus\ states in \eightbe\ \cite{simpson2016}.
The 3.03 MeV \twoplus\ state of \eightbe\ has an on-resonance width of $\Gamma(E_R)= 1513 \pm 15$ 
keV, and thus a mean life of $\tau = \hbar/\Gamma(E_R) = 0.44 \times 10^{-21}$ s 
\cite{Tilley2004}. 
As such, breakup from this state will occur very close to the target-like nucleus. To determine the 
effect on complete fusion, it is 
then necessary to quantitatively understand whether such short mean lives carry a significant 
fraction of excited projectile-like nuclei inside the fusion barrier before breakup occurs, thus 
reducing the suppression of complete fusion due to breakup. 

To address this question, this work presents a 
re-analysis of the extensive sub-barrier breakup measurements of Rafiei \textit{et al.} 
\cite{Rafiei2010b}, 
using a modified version of \platypus\ which incorporates resonance lifetimes. The re-analysis of 
these experimental data is presented in 
Sec. \ref{analysis}.  An improved method has been used to better determine the coincidence 
detection efficiency of the detector array, discussed in Sec. \ref{efficiency}. As a 
result of these changes, a different efficiency correction for the 
detector geometry has resulted, which feeds back into the determination of the breakup function -- 
the probability of breakup along a trajectory with distance of closest approach $R_{min}$ -- 
given as input into \platypus. Model sensitivities to breakup observables and the resultant 
modifications to \platypus\ are discussed in Sec. \ref{platypus}. New below-barrier breakup
functions are derived in Sec. \ref{fcns}. The calculations of above-barrier fractions of 
incomplete fusion are presented in Sec. \ref{FICF}, and the consequences of these calculations for 
the role of breakup in the suppression of complete fusion is discussed in Sec. \ref{conclusions}.

\section{Data analysis \label{analysis}}

Full experimental details for the data analysed here can be found in Ref. 
\cite{Rafiei2010b}, and a brief summary is given here for completeness. Beams 
of \ninebe\ at below-barrier energies were delivered by the 14UD electrostatic accelerator at 
the Australian National 
University Heavy Ion 
Accelerator Facility onto isotopically enriched targets of \sm, \er, \w, \pt, $^{208}$PbS and 
\bi. The breakup of 
\ninebe, whether direct or triggered by neutron stripping, results in  
two coincident $\alpha$ particles. The Breakup Array for Light Nuclei 
(BALiN) was used to detect these coincident fragments. The array is composed of four Double-sided 
Silicon Strip Detectors (DSSDs), each with 16 arcs and 8 sectors, resulting in 512 effective 
pixels 
over the array. Below-barrier ($E/V_B \sim 0.65 - 0.9$) measurements of coincident 
$\alpha-\alpha$ pairs were made, as reported in Ref. \cite{Rafiei2010b}, with the goal of 
extracting breakup 
probabilities as a function of the distance 
of closest approach.  In analyses such as these, the challenge is in
separating coincident breakup events from all other reaction outcomes that result in coincident 
signals in a detector array. Genuine coincident breakup events 
were distinguished from spurious coincidence events (mainly resulting from random coincidences 
between scattered 
projectiles and electronic noise), by selecting the characteristic diagonal bands that appear 
when plotting the energy of one coincident particle ($E_1$) against the energy of the other ($E_2$) 
(see, for example, Fig. 3 of Ref. \cite{Rafiei2010b}). For completeness, in Appendix 
\ref{AppendixA} 
we describe an improved method for removal of spurious coincidence events resulting from cross-talk 
or particles crossing an interstrip partition, which are not removed by $E_1$-$E_2$ gating.

  \begin{figure*}
   \begin{center}
   \includegraphics[width=\textwidth]{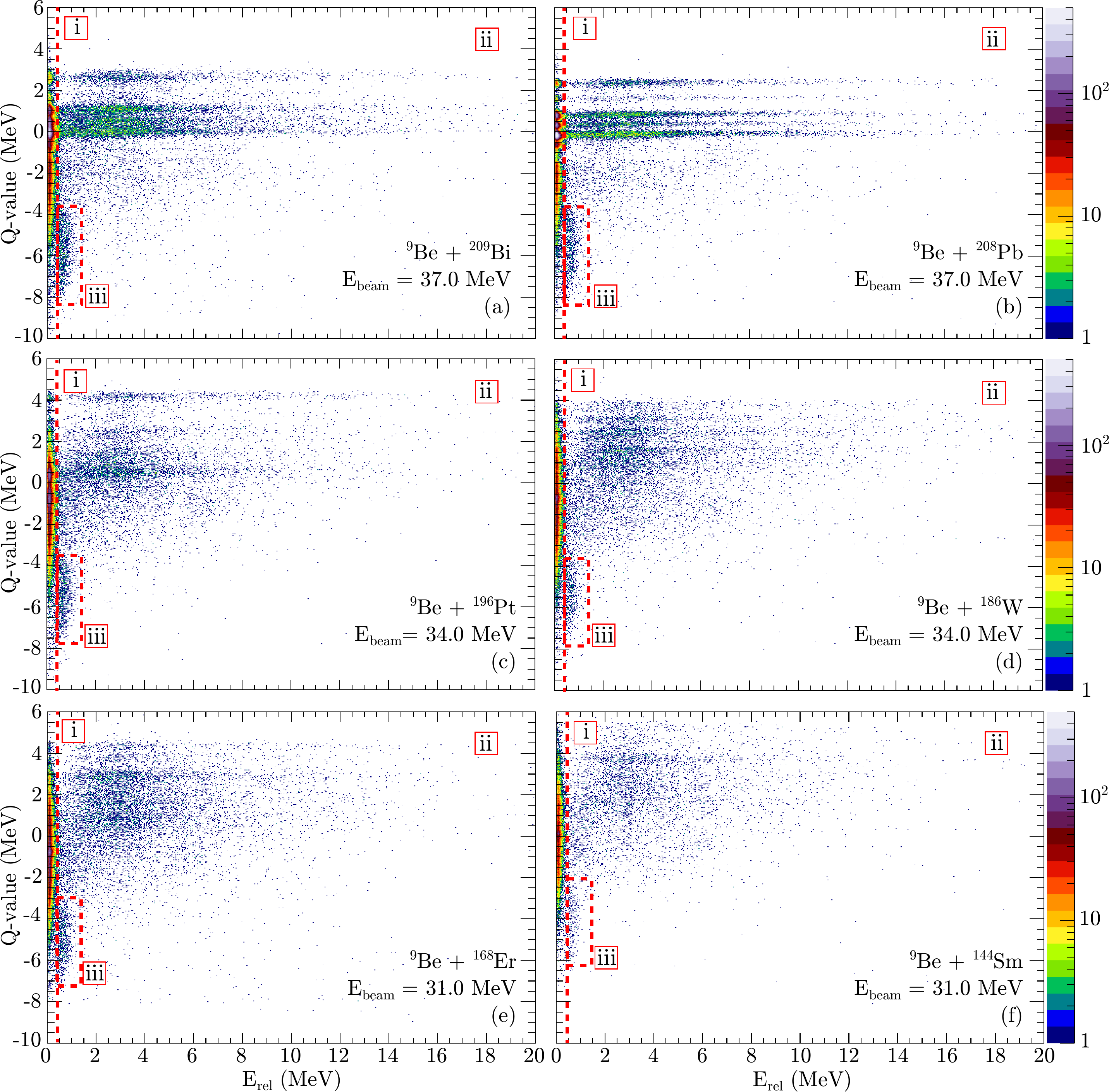}
    \caption{(Colour online) Spectra of the reconstructed \erel\ against Q-value for the 
reactions studied in this re-analysis. Measurements at centre-of mass energies of $E_{c.m.}/V_{B} 
\sim 0.9$ are 
shown. Events arising from the breakup of \eightbe\ from its \zeroplus\ ground state, which 
includes contributions from direct (\ninebe$\rightarrow ^8$Be$_{0^+} + n$) and transfer-triggered 
breakup are shown to the left of the vertical dashed line denoted region \textit{(i)}. Events from 
breakup of \eightbe\ from either the high excitation energy tail of the \zeroplus\ state or the 
\twoplus, $4^{+}$ states lie to the right of the line [region \textit{(ii)}], excepting those 
marked by the dashed box, \textit{(iii)}, which contains direct breakup events from the decay of 
\ninebe\ from its \fivehalf\ state.}
\label{fig:QErel}
    \end{center}
   \end{figure*}

\subsection{Distinguishing near-target and asymptotic breakup}
After the removal of spurious events, the reconstructed spectra of reaction Q-value against 
relative energy of the two coincident breakup fragments, $E_{rel}$, for \ninebe\ + \heavy\ at 
centre of mass energy $E_{c.m.}$ such that $\frac{E_{c.m.}}{V_\textrm{B}} \sim 0.9$ are 
shown in Fig. 
\ref{fig:QErel}. Compared to previous results, at this stage in the re-analysis, the data differ 
mainly in the larger number of events from \eightbe\ ground-state decay, as noted in Appendix 
\ref{AppendixA}. The Q-values are determined by
\begin{equation}
\mathrm{Q} = (E_1 + E_2 + E_\mathrm{recoil}) - E_\mathrm{lab},
\end{equation}
where $E_i$ are the energies of each fragment, corrected for energy loss through the target, 
mylar foil, aluminium layer and silicon dead-layer, $E_\mathrm{lab}$ is the beam energy after 
traversing half the target thickness, and $ 
E_\mathrm{recoil}$ is the energy of the recoiling target-like nucleus, which is determined through 
momentum conservation. As discussed in Ref. \cite{Rafiei2010b}, the distribution of Q-values 
reflects the excitation of the target-like nucleus. The \erel\ distribution is determined 
using the expression
\begin{equation}
E_{rel} = \frac{m_2 E_1 + m_1 E_2 - 2\sqrt{m_1 E_1 m_2 E_2}\cos\theta_{12}}{m_1 + m_2},
\end{equation}
where $\theta_{12}$ is the measured laboratory frame opening angle, given by
\begin{equation}
\cos\theta_{12} = \cos\theta_1\cos\theta_2 + \sin\theta_1\sin\theta_2 \cos(\phi_1-\phi_2),
\end{equation}
\noindent $m_i$ and $E_i$ are the mass and energy of each fragment, and 
($\theta_i,\phi_i$) is the measured scattering angle and azimuthal angle of each 
signal in a coincidence event. The \erel\ distribution reflects the excitation 
of the projectile-like nucleus, modified by post breakup Coulomb interactions of the fragments with 
the target-like nucleus. Events with small relative energy \erel$ \lesssim 180$ keV, labelled 
region 
\textit{(i)}, result from breakup of the \eightbe\ ground state with \erel$ = 92$ keV. The spread 
in 
measured \erel\ up to 180 keV results from the angular size of the detector pixels 
\cite{Luong2013}. This unbound state has a width $\Gamma = 5.57 \pm 0.25$ eV, and therefore mean 
life $\tau = 1.2 \times 10^{-16}$s \cite{Tilley2004}. Due to this long mean life, breakup 
will occur asymptotically far from the target-like 
nucleus, such that the gradient of the Coulomb field accelerates the two fragments in essentially 
the same direction. 

On the other hand, events with large \erel, labelled as region \textit{(ii)}, is associated with 
breakup of \eightbe\ which 
results in high relative energy. One contributor to such events is breakup of \eightbe\ from its 
\twoplus\ resonant state. This state has a large width, $\Gamma = 1513\pm 15$ keV, and 
thus a mean life $4.35 \times 10 ^{-22}$ s \cite{Tilley2004}. \eightbe\ populated in this state
will therefore break up close to the target-like nucleus, where 
fragment-target interactions significantly affect the trajectory of the breakup fragments. It is 
these events that may influence complete and incomplete fusion cross-sections at above-barrier 
energies.

Events resulting from direct breakup of \ninebe\ ($^9$Be$^* \rightarrow \alpha + \alpha + n$) from 
the 2.43 MeV $\frac{5}{2}^-$ state \cite{Tilley2004} are grouped in the region labelled 
\textit{(iii)} in each panel. The spread in Q-values reflects 
the fact that the energy carried by the neutron is not captured by the 
\balin\ array, resulting in an incorrect reconstruction of the \qval\ of this breakup mode. Despite 
missing the neutron, the distribution is relatively sharply peaked in \erel\ ($\sim 
0.6$ MeV) reflecting the long mean life of $\tau \sim 8.4 \times 
10^{-19}$ s ($\Gamma = 0.78 \pm 0.13$ keV) of this state \cite{Tilley2004}. Breakup 
from this state will thus also occur far from the target-like nucleus, giving minimal differential 
acceleration of the $\alpha$ particles following breakup.
 
   \begin{figure}
   \begin{center}
   \includegraphics[width=\columnwidth]{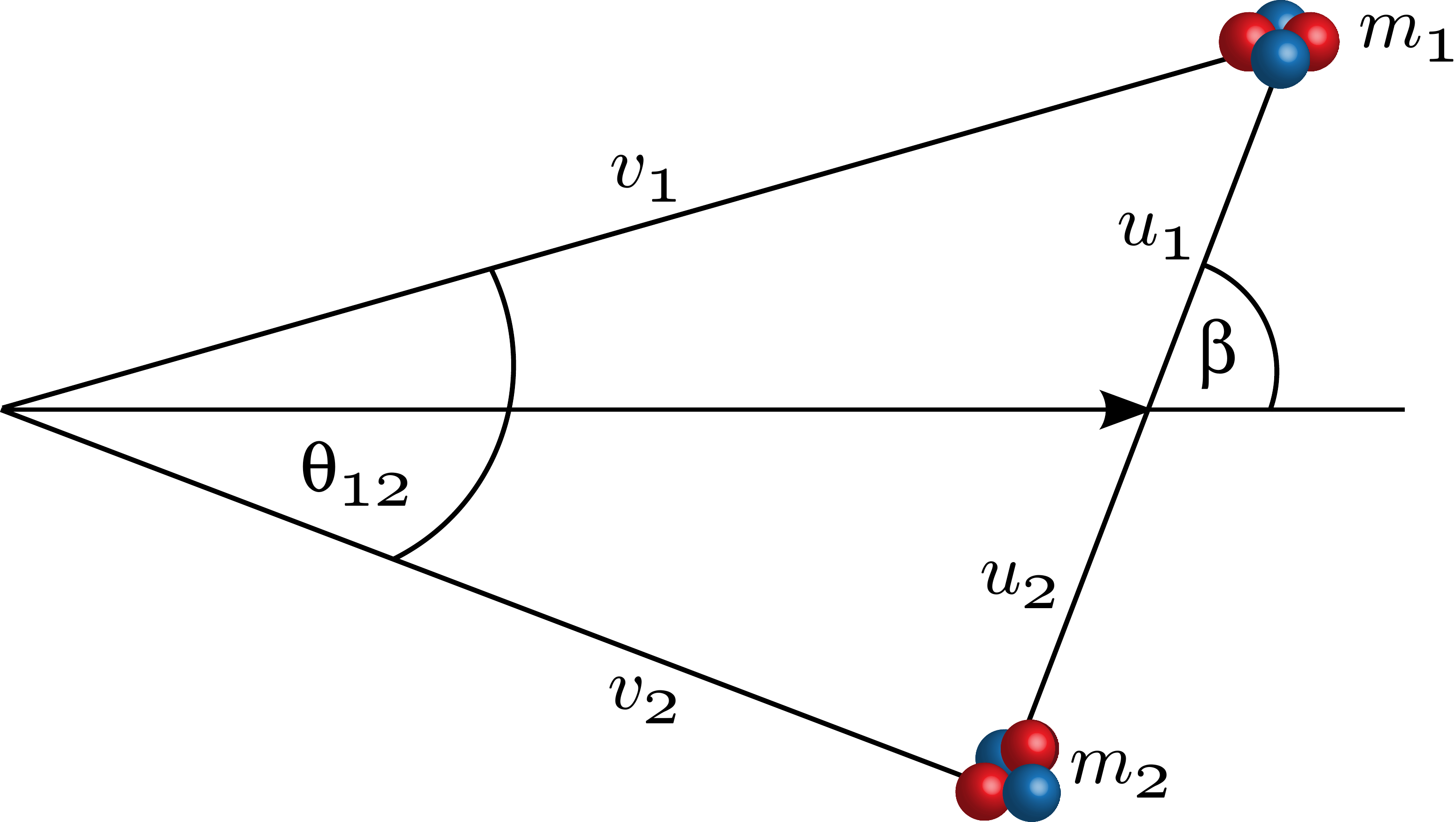}
    \caption{(Color online) Diagram demonstrating the relationship between opening angle 
$\theta_{12}$ and the orientation of the relative momenta of the breakup fragments $\beta$. $v_i$ 
is the laboratory velocity for each fragment with mass $m_i$, and is deduced from their measured 
energy $E_i$. $u_i$ is the velocity of each fragment in their centre of mass frame, deduced from
momentum conservation and their relative energy.    }
    \label{fig:betatheta12}
    \end{center}
   \end{figure}

\begin{figure}
\includegraphics[width=\columnwidth]{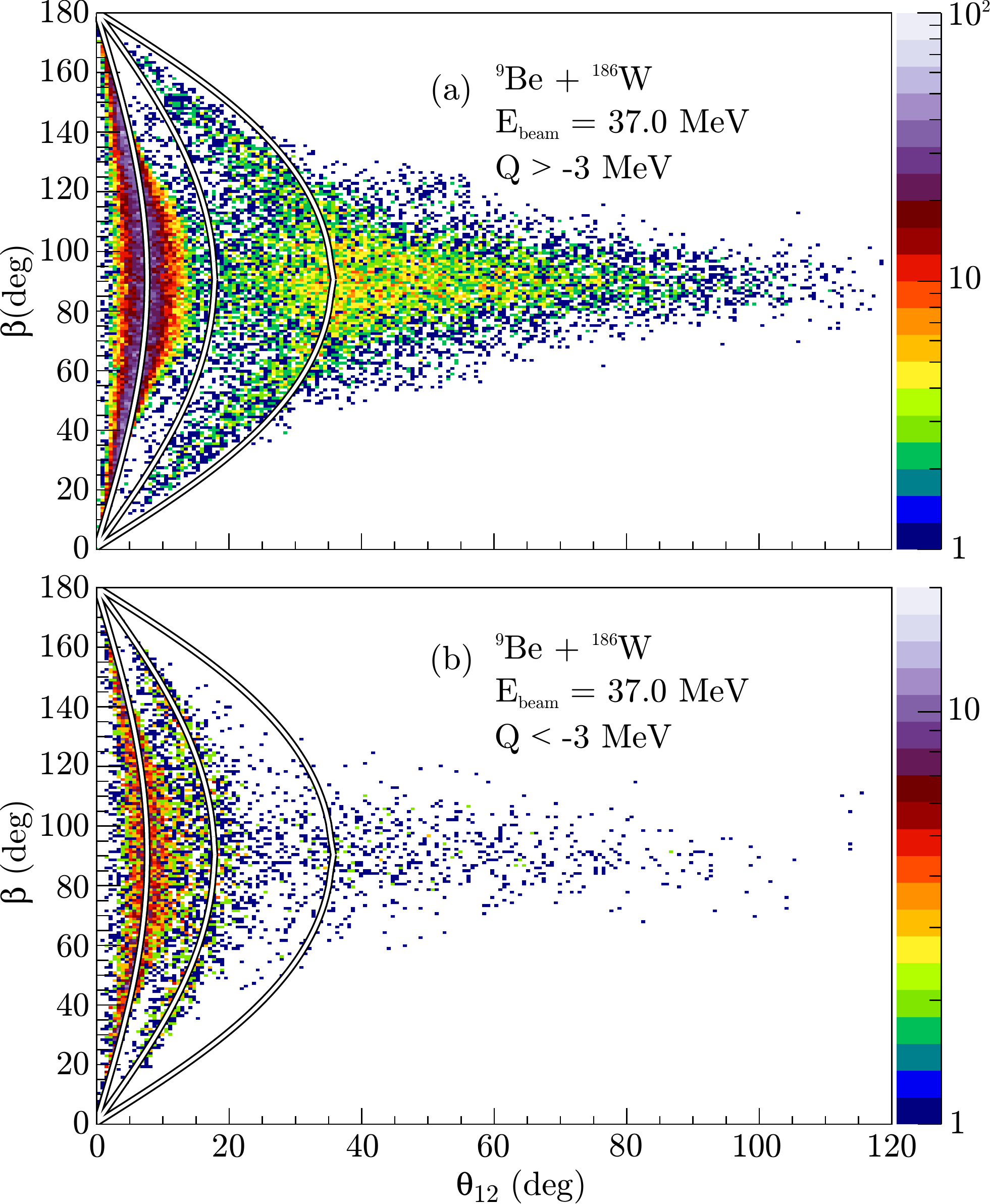}
\caption{(Colour online) Deduced experimental $\theta_{12}-\beta$ distribution for the breakup of 
\eightbe\ 
formed after neutron transfer from \ninebe\ in interactions with \w\ at $E_\mathrm{beam}=37$ MeV.
Panel (a) shows  $Q>-3$ MeV to highlight transfer-triggered breakup, and (b) shows events with 
$Q<-3$ MeV, where the direct \ninebe\ \fivehalf\ curve is more clearly seen. 
Lines indicate $\theta_{12}-\beta$ curves calculated for the asymptotic breakup of (left to right) 
\eightbe\ 
\zeroplus, \ninebe\ 
$\frac{5}{2}^-$ and \eightbe\ \twoplus. Distributions that deviate from these curves are 
a result of breakup that occurs sufficiently close to the target-like nucleus to perturb the final 
trajectories of the breakup fragments. If particles fall into the same pixel of \balin, they do not 
register as coincidence events, resulting in a reduced number of events observed near 
$\beta = 0^\circ$ and $180^\circ$.}
\label{fig:Wthetabeta}  
\end{figure}

Independent of expectations based on the known mean lives of resonant states, deduced from their 
widths, it is 
possible to experimentally separate breakup close to the target nucleus from breakup 
(asymptotically) far away by examining the energy and angular 
correlations of the resulting fragments \cite{simpson2016,McIntosh2007}. When 
breakup occurs asymptotically, which is also associated with a well defined excitation energy 
$E_{x}$ of the projectile-like nucleus, the laboratory opening angle between the two 
fragments, \thetaonetwo, and the 
orientation of the relative momentum of the breakup fragments, $\beta$, in 
their centre of mass 
frame are related. These quantities,  \thetaonetwo\ and $\beta$, are shown in 
Fig. 
\ref{fig:betatheta12}, which can be used to obtain the relationship: 
\begin{equation}
\sin\beta = \frac{v_1 v_2 \sin\theta_{12}}{(v_2^2 u_1^2 + v_1^2 u_2^2 + 
2u_1u_2v_1v_2\cos\theta_{12})^{1/2}}.
\label{eqn:beta}
\end{equation}
Here, $v_i$ is the laboratory velocity for each fragment, deduced from their measured 
energy $E_i$, 
and $u_i$ is the velocity of each fragment in their centre of mass frame, deduced from
momentum conservation and their relative energy $E_\mathrm{rel} = \frac{1}{2}\mu_{12}(u_1 + 
u_2)^2$, $\mu_{12}=\frac{m_1 m_2}{m_1+m_2}$. The 
$\theta_{12} - \beta$  distributions, reconstructed from the measured data for \ninebe + \w\ 
at $E_{beam} = 37.0$ MeV are shown in Fig. \ref{fig:Wthetabeta} for $Q > -3$ MeV (panel a), 
where transfer-triggered breakup is dominant, and for $ Q <-3$ MeV (panel b). The latter 
includes contributions from direct breakup, which are those shown in region \textit{(iii)} of 
Fig. \ref{fig:QErel}(d) for the same system at $E_{beam} = 34.0$ MeV. The lines overlaid on the 
data in Fig. \ref{fig:Wthetabeta} correspond to 
calculations using Eqn. (\ref{eqn:beta}) for $E_{x}$ corresponding to breakup from (from 
left to right) \eightbe\ 
\zeroplus, $E_x = 92$ keV, \ninebe\ \fivehalf, $E_x = 600$ keV [region \textit{(iii)} in Fig. 
\ref{fig:QErel}], and \eightbe\ \twoplus, $E_x = 3.03$ MeV. As can be seen in the figure, 
bands with excellent correspondence to the calculations for the asymptotic breakup of 
\eightbe\ \zeroplus\ and 
\ninebe\ \fivehalf\ are 
present 
in the experimental \thetabeta\ distribution, confirming the interpretation that these events 
correspond to breakup asymptotically far from the target-like nucleus. However, as can be seen in 
Fig. \ref{fig:Wthetabeta}(a), the calculation assuming
asymptotic breakup of \eightbe\ \twoplus\ does not match the data well. This can be explained as 
a result of breakup occurring close to the target-like nucleus. When this occurs, the initial 
kinetic energy of the fragments is small, and their energies are stored in the 
fragment-target potential. As a result, there is an increased probability for $E_{1}\sim E_{2}$ 
and thus of deduced values of $\beta \sim 90$\textdegree\ for breakup into identical fragments
\cite{simpson2016}.
Therefore, without making any assumption of the state that is populated, the concentration of 
events around 
$\beta \sim 90$\textdegree, indicates breakup close to the target-nucleus. Thus it is these 
events that may influence complete fusion cross-sections. The extraction of breakup probabilities 
for these events is the subject of Sec. \ref{efficiency} and \ref{fcns}.

\subsection{Cross-section normalisation}
In order to extract breakup probabilities, the array was partitioned into $5^\circ$ bins 
covering laboratory angles from $\theta = 130^\circ$ to $165^\circ$. The yield of breakup events 
in each bin must be normalised to the yield of Rutherford scattering. Elastic events for 
normalisation were extracted 
from a $\theta$ bin of the \balin\ array from 
124\textdegree\ -- 
127\textdegree, where the elastic yield is pure Rutherford for deep sub-barrier measurements. At 
higher energies, the yield was corrected by up to 11$\%$, determined from optical model 
calculations, described in Appendix \ref{AppendixB}. Recent
precision measurements of the spatial positioning of the \balin\ detectors have resulted in slight 
changes in the location of the array relative to the beam axis. This has resulted a $9 \pm 1\%$ 
decrease in the number of elastic particles assigned to
the 124\textdegree\ -- 127\textdegree\ bin for each measurement compared to those reported in Ref. 
\cite{Rafiei2010b}.

\section{Improved method for coincidence efficiency determination \label{efficiency}}
In order to determine absolute breakup probabilities, the coincidence efficiency $\epsilon$ of 
\balin\ had to be determined. In this work, a new two-step approach to efficiency determination was 
implemented with minimal reliance on simulated breakup distributions. The first step was to 
calculate the geometric coincidence efficiency of the \balin\
array as a function of \thetaonetwo\ and breakup pseudo-angle $\theta_{^8\mathrm{Be}}$ (described 
below). A Monte-Carlo simulation was used to obtain these geometric coincidence efficiencies. 
\platypus\ was used as the Monte-Carlo simulator, but the efficiencies derived in this step were 
model-independent. However, the geometric coincidence efficiencies did not account 
for the events that fall outside of the detector acceptance in ($\theta_{^8\mathrm{Be}}$, 
$\theta_{12}$). The second step in the efficiency determination was to simulate the 
total distribution of fragments to correct for those events with $\theta_{12}$ that 
fall outside the detector acceptance for each $\theta_{^8\mathrm{Be}}$. This correction was small 
-- the events comprised $\sim7\%$ of the total yield in the $\theta_{^8\mathrm{Be}}$ acceptance of 
the detector. These simulations were done using a version of \platypus\ which incorporated the 
modifications discussed in Sec. \ref{platypus}. Full details of the efficiency determination is 
described in Appendix \ref{Appendix:Efficiencies}, and a comparison of the efficiencies calculated 
in this work to those of Ref. \cite{Rafiei2010b} is presented in Appendix 
\ref{Appendix:Comparison}.

The breakup pseudo-angle is also needed in order to extract breakup functions from 
coincidence data. When a reaction produces only one nucleus related to the lighter collision 
partner in the outgoing trajectory, the angular 
distribution and distance of closest approach of the projectile and target nuclei may be estimated
from the measured scattering angle in a straightforward manner. In a breakup reaction producing 
pairs of particles which will have different angles $\theta$ and $\phi$, an appropriate way to 
extract the breakup function is by use of
$\theta_{^8\mathrm{Be}}$, which can be interpreted as the reconstructed 
scattering angle of the \eightbe\ had it not broken up. This is related to the deduced recoil angle 
of the target-like nucleus $\theta_\mathrm{recoil}$. The latter is already used to calculate the 
kinetic energy of the recoiling target-like nucleus and thus the Q-value of the breakup reactions.  
$\theta_\mathrm{recoil}$ is determined from the momenta of the measured breakup fragments using 
momentum conservation, and $\theta_{^8\mathrm{Be}}$ is given by 
\begin{equation}
\tan \theta_{^8\mathrm{Be}} = \frac{\sin 2\theta_{recoil}}{M_p/M_t -
\cos2\theta_{recoil}}, 
\end{equation}
where $M_p$ is the mass of the projectile-like nucleus and $M_t$ the mass of the 
target-like nucleus.

\section{Classical trajectory simulations\label{platypus}}

The ultimate aim of this work is to understand the contribution that transfer triggered breakup 
makes to the suppression of complete fusion at energies above the barrier. By making below-barrier 
measurements of no-capture breakup probabilities and relating these probabilities to above-barrier 
CF and ICF cross-sections, it is possible to determine the contributions of 
breakup to suppression of CF, and to cross sections for ICF products. However, to achieve this, a 
reliable simulation of post-breakup trajectories of the fragments is required. This is for two 
reasons: firstly, to extract the below-barrier near-target breakup probabilities 
from experimental results, and secondly, to take these experimentally determined breakup 
probabilities and make predictions of CF and ICF at above-barrier energies.

As no fully quantum mechanical model of transfer induced breakup exists yet, classical simulations 
have been performed. Clearly, it is important that a classical model captures the key physics of 
the breakup processes. Namely, (a) the locations of the transfer reactions, (b) the properties of 
the intermediate nucleus populated after transfer, and (c) the subsequent decay and 
post-breakup acceleration of the fragments. The acceleration of the fragments after breakup has the 
capacity to change their relative energy, and is the classical analogue of continuum-continuum 
couplings in quantum mechanical models. The classical dynamical breakup 
code \platypus\ \cite{Diaz-Torres2007,Diaz-Torres2011}, with modifications described below, 
provides an appropriate platform for these calculations. \platypus\ is a three-body 
classical trajectory model with stochastic breakup that enables calculations of breakup observables 
as well as incomplete and complete fusion cross-sections. It considers a target and a 
weakly-bound pseudo-projectile (here, \eightbe) that initially follow Rutherford trajectories. 
Breakup probabilities and locations are  stochastically sampled from an experimentally determined 
breakup function $P(R_\mathrm{min})$. At the point of breakup, the properties of the fragments 
(excitation energy $E_x$, separation, orientation) are stochastically sampled before propagating in 
the fragment-fragment and fragment-target fields. Several significant modifications to \platypus\ 
have been made to more accurately capture the details of breakup dynamics, as described below. 

\subsection{Incorporating excitation energies and lifetimes of resonant states of 
the projectile-like nucleus\label{exandlife}}

In order to include the known low-energy structure of \eightbe, 
modifications to \platypus\ were made to model the resonant states in \eightbe. The energy 
and angular distribution of breakup fragments produced after the decay of a projectile-like nucleus 
populated in transfer reactions depends critically on  (i) the excitation of the projectile-like 
nucleus that breaks up, and (ii) the location of breakup with respect to 
the target-like nucleus, which is in turn sensitive to the lifetime of the projectile-like nucleus 
after the point of transfer. In the previous versions of \platypus, the excitation of the 
projectile-like nucleus was given as a range from $E_\mathrm{min}$ to $E_\mathrm{max}$ with 
either a flat or exponentially decreasing distribution \cite{Diaz-Torres2011}. Although lifetimes 
were not treated explicitly, breakup fragments would take some time to propagate from their 
assumed initial Gaussian distribution of separations to beyond their mutual barrier radius
\cite{Diaz-Torres2007}. This effective lifetime is sensitive to the fragment-fragment potential. 

\begin{figure}
\begin{center}
\includegraphics[width=\columnwidth]{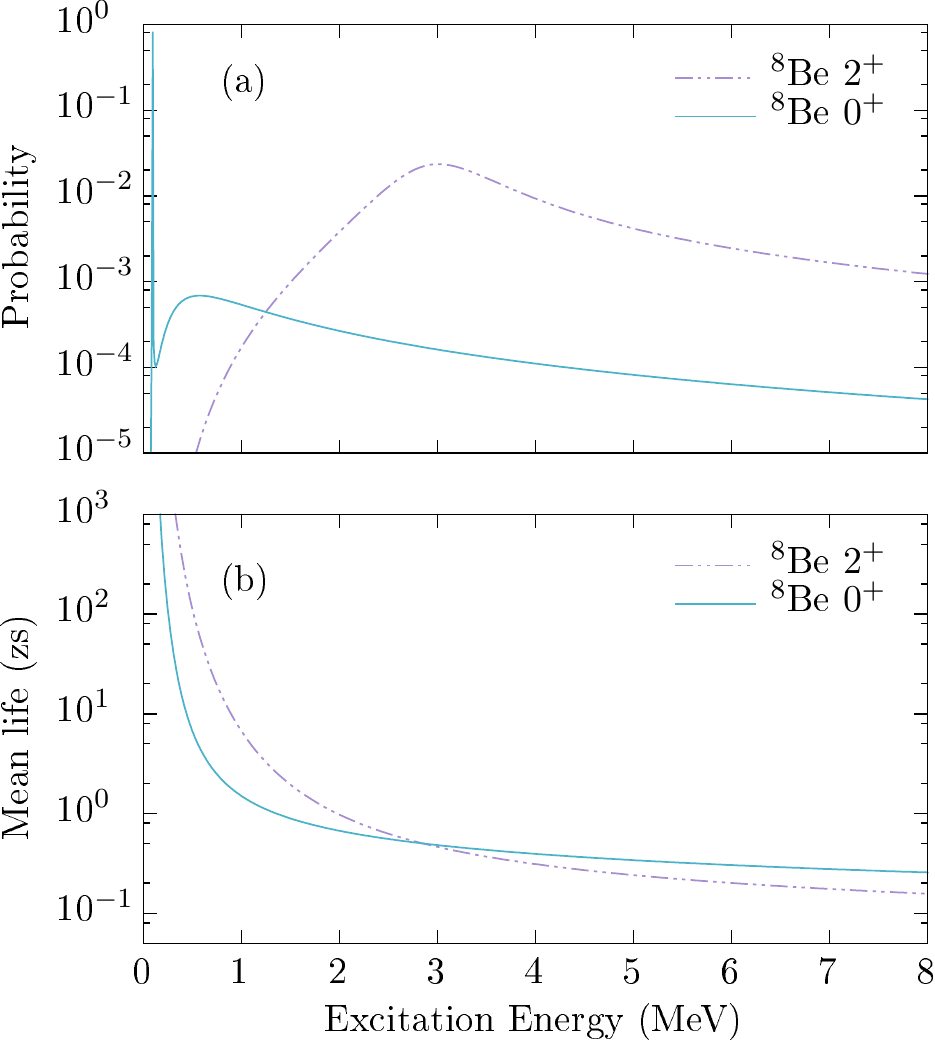}
 \caption{(Colour online) Excitation energy probability distribution (a) and excitation energy 
dependant mean life (b) for \zeroplus\ (solid line) and \twoplus\ (dashed line) states in 
\eightbe\ used as input in the modified version of \platypus that explicitly takes into account 
resonance excitation energies and lifetimes.} 
 \label{fig:distroEx}
 \end{center}
\end{figure}

As 
will be demonstrated 
below, the population of \eightbe\ in the reactions studied in this work can be well described as a 
combination of $0^+$ ground-state and first excited $2^+$ state. Thus, the simulated excitation 
energy and lifetime distributions of \eightbe\ should correspond to the width of 
these states. Modifications to \platypus\ were made such that excitation energies sampled from 
realistic distributions of excitation energy have a corresponding mean life associated with 
each excitation energy. The excitation energy probability distributions were 
calculated from the one-state, one-channel limit of R-matrix theory \cite{Barker1962,Barker1988}. 
The corresponding mean life was estimated using $\tau(E_{x})= \hbar/\Gamma_\ell(E_{x})$, where 
$\Gamma_\ell(E_{x})$ is the energy-dependant resonance width. This has been recently described in 
Ref. \cite{kalkal2016}, where excitation energy probability distributions were calculated for 
$^{6,7}$Li. Shown in Fig. \ref{fig:distroEx} are the resulting excitation energy probability 
distributions (a) and excitation energy dependent mean lives (b) for the \eightbe\ $0^+$ and $2^+$ 
states used in the \platypus\ calculations in this work.

Including these probability distributions of excitation energy and associated mean-life, the 
distribution of decay (breakup) times of short-lived resonance states are now modelled 
explicitly in \platypus. The first step is randomly choosing a ``transfer 
radius'', $R_\mathrm{Tr}$, according to the breakup function as originally done. Then a classically 
allowed excitation energy $E_{x}$ (with corresponding mean life $\tau$) is chosen from the 
distribution of excitation energies as shown in Fig. \ref{fig:distroEx}. 
The projectile then propagates along its trajectory for some time $t$, sampled from the exponential 
distribution of times expected from the 
mean life, $e^{-t/\tau}$, before breaking  up into two fragments with relative energy 
corresponding to 
$E_{x}$. The fragments are initially placed at a separation radius corresponding to the peak of 
their mutual barrier. Breakup is thus defined to occur when the two 
fragments pass their mutual barrier. Crucially, \eightbe\ produced by transfer before the distance 
of closest approach may pass the turning point and begin to recede from the target before breaking 
up. 

This explicit 
handling of excitation energies and mean lives gives a more physically 
realistic (though still phenomenological) distribution of (i) breakup fragment energy and (ii) the 
time taken between transfer and breakup, and thus positions along the 
trajectories. The latter modification in particular removes sensitivity to 
the fragment-fragment potential. In addition, these modifications allow long-lived states, such as 
the \eightbe\ ground state, to be simulated with \platypus\ rather than requiring an additional 
simulation with a different code \cite{Rafiei2010b}. Further, requiring that the distribution of 
excitation energies used in \platypus\ 
be determined by the known resonance properties of \eightbe\ removes 
this quantity as a parameter in the model and, as discussed in Sec. \ref{FICF}, has a significant 
effect on CF and ICF predictions.  

\subsection{Incorporating effects of excitation of target-like nuclei}
As can be seen by the spread of Q-values in Fig. \ref{fig:QErel}, the target-like nucleus is 
populated with a large range of 
excitations (up to $\sim 8$ MeV) in these reactions. Trivially, as the excitation energy of the 
target-like nucleus increases, the energy available for the excitation of the projectile-like 
nucleus 
decreases. This results in a decrease in $E_\mathrm{rel}$ (as can also be seen in Fig. 
\ref{fig:QErel}), and thus a decrease in average opening angle $\theta_{12}$. Therefore, the 
fidelity of the reproduction of experimental results in \platypus\ is also dependent on the 
distribution of target-like excitations.

\platypus, being a classical model, has radii around the 
classical turning point where transfer is classically forbidden due to energy conservation.
The size of this region depends on the beam energy, angular momentum and the excitations of the 
projectile-like and target-like nuclei. The latter was not incorporated in the original 
version of \platypus, which was thus modified to include the excitation energy distribution of 
the target-like nucleus, obtained from the experimentally determined Q-value 
distribution. As a result, the \platypus\ simulation now reflects both the excited states of 
the target-like nucleus and the probability of populating those states in the neutron transfer 
reactions studied in this work. To model the excitation energy, at $R_\mathrm{Tr}$ an equivalent 
amount of kinetic energy is deducted from the projectile-like nucleus such that the direction of the 
relative velocity of the system is maintained. 

\subsection{Modifications to the local breakup function}

The aim of these below-barrier measurements of breakup is to determine the breakup probabilities 
$P$ as a function of $R_{min}$, the distance of closest approach on a Coulomb trajectory. The 
experimental data were fitted with the functional form 
\begin{equation}
P(R_\mathrm{min}) = e^{\mu R_\mathrm{min} + \nu},
\label{breakupfunction}
\end{equation}
where $\mu$ and $\nu$ are the (logarithmic) slope and intercept of the function, 
respectively. This function is interpreted as the integral of the local reaction probability 
$\mathcal{P}(R)$ along the classical orbit of the projectile,
\begin{equation}
P(R_\mathrm{min}) = 2 \int_{R_{min}}^{\infty} \mathcal{P}(R) dR.
\end{equation}
$\mathcal{P}(R)$ is a function of the projectile-target separation $R$, and 
$\mathcal{P}(R)dR$ gives the reaction probability between $R$ and $R+dR$.
The factor of two reflects the initial assumption that taking breakup to be instantaneous, it can 
occur with equal probability on the ingoing and outgoing trajectories. With the incorporation of 
resonance lifetimes, the local probability must now be interpreted as that for the trigger event 
for breakup, in this case transfer. At above-barrier energies, when using \platypus\ to estimate  
$\sigma_\mathrm{ICF}$, the distance of closest approach is inside 
the barrier radius, thus only the transfer probabilities on the ingoing trajectory should 
included. This change by a factor of two has been taken into account in the modified \platypus\ 
calculations of $\sigma_\mathrm{ICF}$, resulting in a decrease in 
contributions to $\sigma_\mathrm{ICF}$ from trajectories with angles within the grazing angle by 
approximately a factor of two. 

The distribution of transfer positions along the projectile-target trajectory has also been 
modified. In the original \platypus, when determining the probability along 
the trajectory it is assumed that since
\begin{equation}
 2 \int_{R_{min}}^{\infty} \mathcal{P}(R) dR =  e^{\mu R_{min} + \nu},
\end{equation}
the local probability must then have the form \cite{Diaz-Torres2007}:
\begin{equation}
\mathcal{P}(R)\propto e^{\mu R}.
\end{equation}
However, this neglects the fact that interacting nuclei spend 
more time near the distance of closest approach than at other distances. As a result 
$dP(R_{min})/dt$ goes to zero at the point of closest approach, as illustrated in Appendix 
\ref{appendixPBU} for a classical Coulomb trajectory.

Instead, we assign each time step on a particular projectile trajectory a relative probability 
assuming a local (transfer) probability $\mathcal{\widetilde{P}}(t) \propto e^{\mu R(t)}$ and 
normalise the full trajectory such that 
\begin{equation}
P(R_{min}) = \int^\infty_{-\infty} \mathcal{\widetilde{P}} 
(t)  dt.
\end{equation}
The local probability is then peaked at the distance of closest approach, which is physically more 
reasonable.

\subsection{Comparison with experimental data \label{sec:orientation}}

\begin{figure*}
\includegraphics[width=\textwidth]{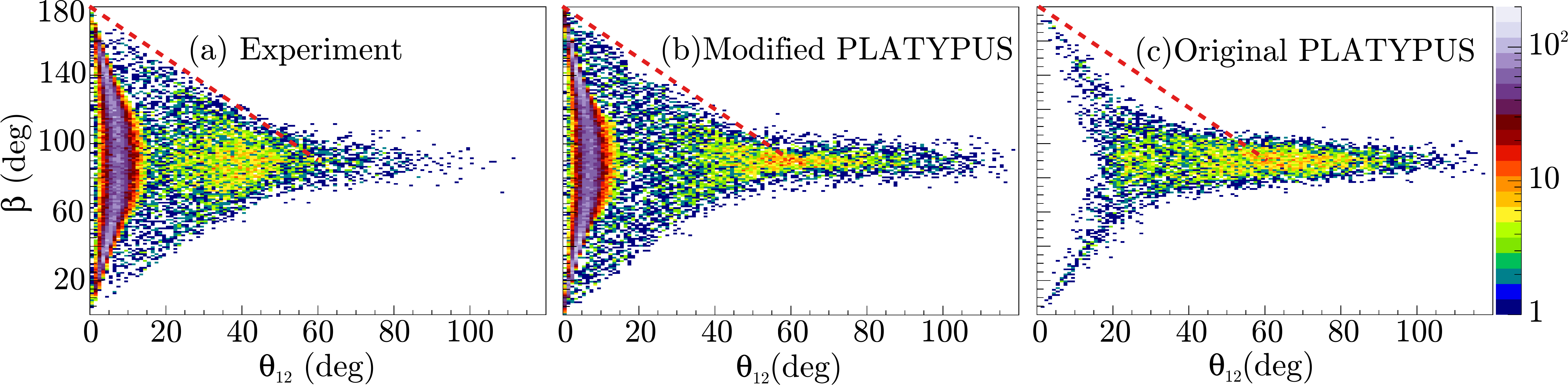}
\caption{(Colour online) (a) Measured $\theta_{12}-\beta$ distribution for the breakup of 
\eightbe\ formed following neutron transfer in interactions of \ninebe\ with \bi\ at 
$E_{beam} 
= 34.0$ MeV. (b) The corresponding 
modified \platypus\ 
simulation, 
which includes contribution from \eightbe\ \zeroplus\ and \twoplus\ resonances. (c) The 
corresponding unmodified \platypus\ simulation, with $0.95\leq E_x\leq4$ MeV, approximating the 
\eightbe\ \twoplus\ resonance only. The red diagonal line provides a reference to quantify the 
differences between the observables for the \twoplus\ resonance.}
\label{fig:thetabeta}  
\end{figure*}

The accuracy of the \platypus\ simulations was assessed by comparing them with the experimentally 
measured $\theta_{12}-\beta$ distributions. The $\theta_{12}-\beta$ distributions are a good test 
as they are sensitive to the effect of fragment-target interactions, and 
therefore to the position and energetics of breakup \cite{simpson2016}. The experimental  
$\theta_{12}-\beta$ 
distribution for the  breakup of \eightbe\ formed following neutron transfer in collisions of 
\ninebe\ with \bi\, is shown in Fig \ref{fig:thetabeta}(a). It is compared with modified and 
unmodified \platypus\ simulations in Fig. \ref{fig:thetabeta}(b) and (c), respectively.  As the 
original \platypus\ does not simulate long-lived states, the $0^+$ state seen in the intense purple 
band at small \thetaonetwo\ in Fig. \ref{fig:thetabeta}(a) has not been included. In the modified 
\platypus\ simulation, both $0^+$ and $2^+$ resonances have been simulated, and the distributions 
are combined to produce the same ratio of breakup events that populate the \erel$= 92 $ keV $0^+$ 
peak to the total number of events as seen in the experimental data. 

As discussed in Sec. 
\ref{analysis}, the effect of fragment-target Coulomb interactions results in  deviations in the 
$\theta_{12}-\beta$ distribution from that expected for asymptotic breakup (calculated using 
Eqn. \ref{eqn:beta}). The modified version of \platypus\ well reproduces the \zeroplus\ peak, and 
reproduces the high \thetaonetwo\ component better than the unmodified model (in particular 
events below the diagonal red dashed line, which is drawn to guide the eye). However, the 
simulation contains a higher intensity of events with \thetaonetwo\ $\gtrsim$\ 60\textdegree\ and 
$\beta \sim 90$\textdegree. This means that too many breakup events result in coincident fragments 
with similar energies and large opening angles. This discrepancy could be ameliorated by considering 
the effect of the projectile-target potential in producing a preferential orientation for \eightbe\ 
relative to the target, as has been previously explored for the direct breakup of $^7$Li  
\cite{mason1992}. However, without a satisfactory method for reliably parameterising orientation 
effects, they are neglected, and all breakup is assumed to occur 
isotropically in the rest frame of \eightbe. Nevertheless, these simulations demonstrate 
that the population of \eightbe\ in the reactions studied in this work can be reasonably well 
described as a combination of \zeroplus\ ground-state and \twoplus\ first excited state. Further, 
the modifications to (i) better model the projectile-like nucleus in resonant states with 
explicitly included mean lives, (ii) model reactions that result in excitation of the target-like 
nucleus, and (iii) better distribute the  transfer probability along the projectile-target 
trajectory provides a more physically realistic, though still phenomenological, model of breakup 
following transfer.

\section{Near-target breakup probabilities\label{fcns}}

The breakup probability is defined for each $\theta_{^8\mathrm{Be}}\sim 5$\textdegree\ bin as the 
ratio between the 
breakup cross-section, determined from the yield of breakup fragments with the reconstructed angle 
of 
the unbroken projectile falling in $\theta_{^8\mathrm{Be}}$, and the Rutherford scattering 
cross-section for each $ \theta_{^8\mathrm{Be}}$ bin,
 \begin{equation}
 P(\theta_{^8\mathrm{Be}}) = 
 \frac{(\frac{d\sigma}{d\Omega})_\mathrm{BU}(\theta_{^8\mathrm{Be}})}{(\frac{d\sigma}{d\Omega}
 )_\mathrm{Ruth} (\theta_{^8\mathrm{Be}})}.
 \end{equation}

The breakup pseudo-angle maps to a distance of closest approach of the target and 
unbroken projectile $R_\mathrm{min}$, neglecting the nuclear potential at these sub-barrier 
energies, according to
\begin{equation}
R_\mathrm{min} = \frac{Z_1Z_2 e^2}{2
E_{c.m.}}\left(1+\frac{1}{\sin\frac{\theta_{^8\mathrm{Be}}}{2}}\right).
\end{equation}
\begin{figure}
\begin{center}
\includegraphics[width=\columnwidth]{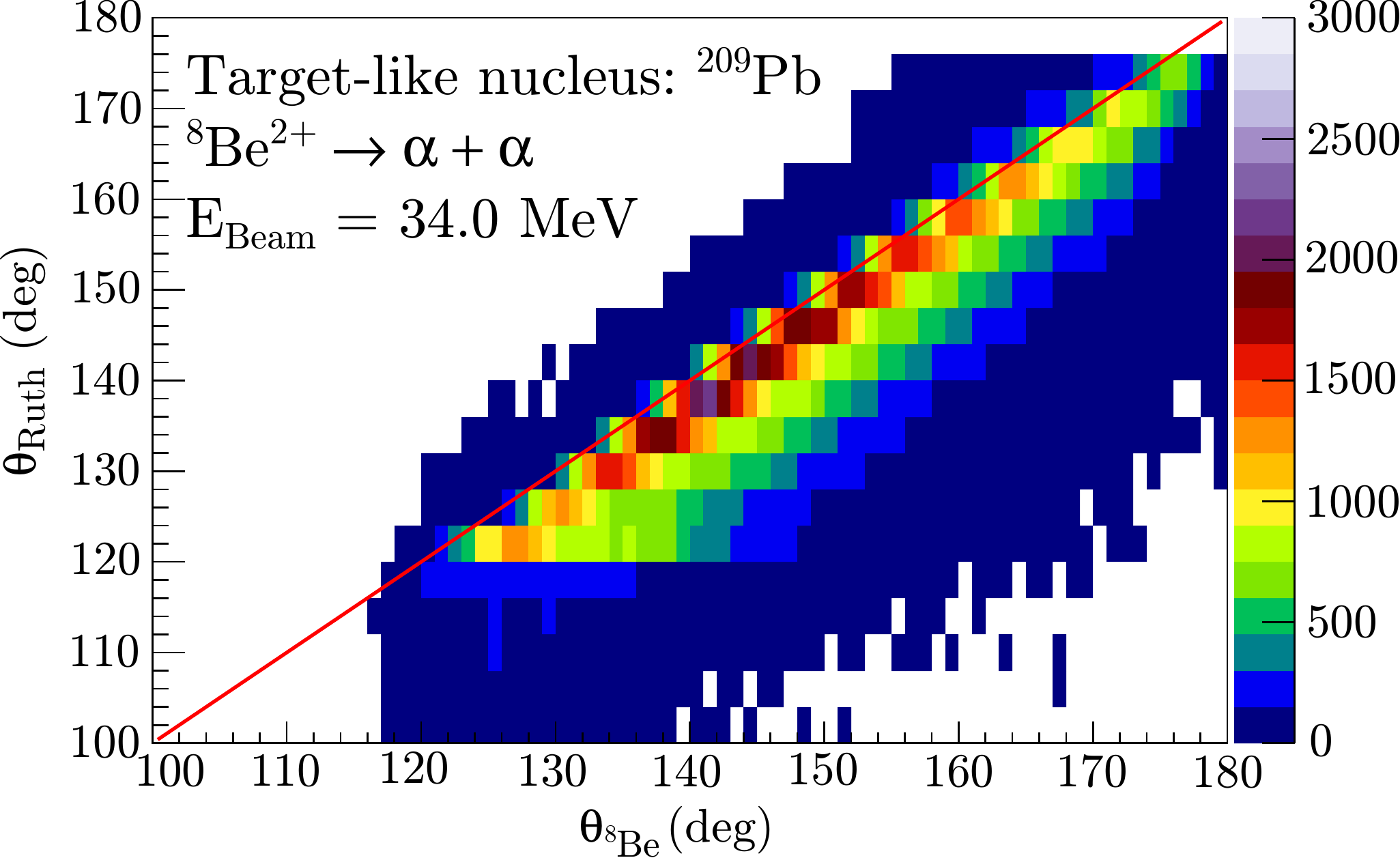}
 \caption{(Colour online) \platypus\ simulation for 
\eightbe$_{2^+}$ + $^{209}$Pb $\rightarrow \alpha + 
\alpha$ + $^{209}$Pb at $E_{beam } =34.0$ MeV, for events that are captured by \balin, 
demonstrating the relatively small difference between 
the Rutherford scattering angle of the \eightbe\ pseudo-projectile $\theta_{\mathrm{Ruth}}$ and 
the angle $\theta_{^8\mathrm{Be}}$ that is reconstructed from the captured $\alpha$ particles.}
 \label{fig:thetaerr}
 \end{center}
\end{figure}
This definition of $R_\mathrm{min}$ implicitly assumes that the reconstructed scattering angle of 
the unbroken projectile-like nucleus is close to the Rutherford angle of the incoming projectile, 
that is, $\theta_{^8\mathrm{Be}} \approx 
\theta_\mathrm{Rutherford}$. This assumption can be tested using \platypus\ simulations. Shown in 
Fig. \ref{fig:thetaerr} is the Rutherford scattering angle of the pseudo-projectile derived from 
the incident trajectory, $\theta_{\mathrm{Ruth}}$, plotted against the reconstructed breakup 
pseudo-angle, for \eightbe$_{2^+}$ + $^{207}$Pb $\rightarrow \alpha + \alpha$ at $E_{beam } 
=34.0$ MeV. In the 
determination of the breakup functions, discussed in Sec. \ref{fcns}, these deviations were treated 
as a correction to $\theta_{^8\mathrm{Be}}$, and for each  $\theta_{^8\mathrm{Be}}$ bin, the 
average 
discrepancy between the Rutherford and reconstructed angles was subtracted from 
$\theta_{^8\mathrm{Be}}$. This correction was larger for breakup that occurs close to the 
target-like nucleus, and was $Z$ dependent, varying from $\sim 1$\textdegree\ for \ninebe + 
\sm, to $\sim 
6$\textdegree\ for \ninebe + \bi. As such, these discrepancies are likely due to trajectories that 
are perturbed by proximity to the high $Z$ target-like nucleus.

With the corrected angle $\theta_{^8\mathrm{Be}}$ transformed to $R_\mathrm{min}$, 
breakup functions may be determined experimentally from the  
ratio of efficiency corrected breakup yield to the elastic yield in each $\theta_{^8\mathrm{Be}}$ 
bin:
 \begin{equation}
 P(\theta_{^8\mathrm{Be}}) = 
\frac{N_\textrm{BU}(\theta_{^8\mathrm{Be}})}{N_\textrm{Ruth} (\theta_\mathrm{Ruth})}.
 \end{equation}
Here $N_{BU}(\theta_{^8\mathrm{Be}})$ is the yield of near-target breakup events 
corrected for efficiency $\epsilon(\theta_{12},\theta_{^8\mathrm{Be}})$, and 
$N_\textrm{Ruth} (\theta_\mathrm{Ruth})$ the calculated Rutherford yield in a given 
$\theta_\mathrm{Ruth}$ bin. Details of the determination of 
$N_\textrm{Ruth}(\theta_\mathrm{Ruth})$ are given in Appendix \ref{AppendixB}.

\begin{figure}
\begin{center}
\includegraphics[width=\columnwidth]{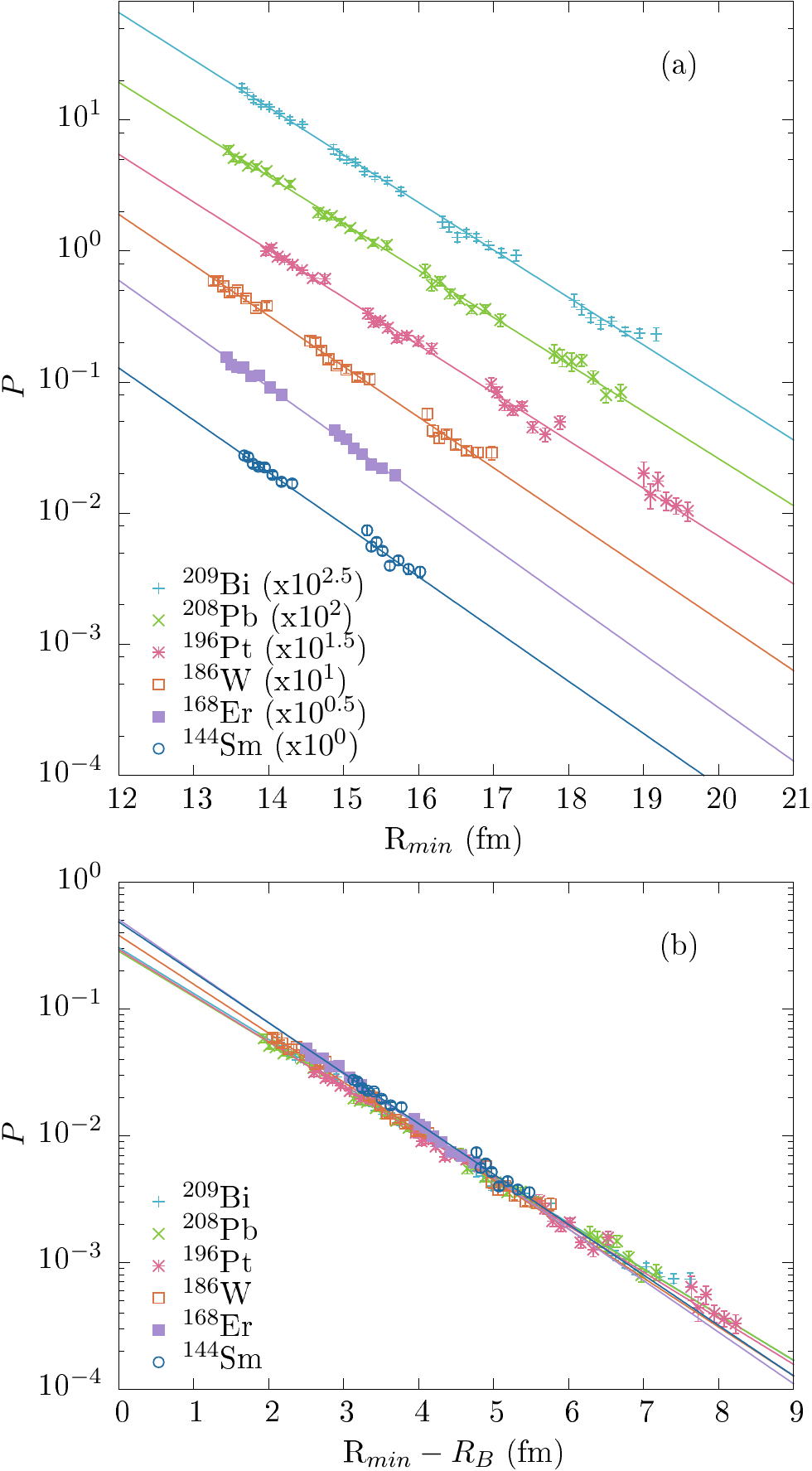}
 \caption{(Colour online) Measured near-target (region $ii$ of Fig. \ref{fig:QErel}) breakup 
probabilities for the breakup of \eightbe\ 
formed following neutron transfer in reactions of \ninebe\  with \heavy\ at energies below the 
barrier (a) as a function of 
the separation of the centres of the nuclei, where probability values have 
been offset for clarity, as indicated in the legend, and (b) as a function of distance from the 
projectile-target barrier. Lines represent least-square fits with Eqn. \ref{breakupfunction}. 
Errors in $P$ are statistical, and for the most part, are smaller than the symbol size.}
 \label{fig:PBU}
 \end{center}
\end{figure}

The resulting probabilities of near-target breakup are 
shown in Fig. \ref{fig:PBU}(a). Each group of points in $R_\mathrm{min}$ represent 
measurements in $5$\textdegree\ $\theta_{^8\mathrm{Be}}$ bins with different $E_\mathrm{beam}$. 
A least-squares fit using Eqn. \ref{breakupfunction} to the experimental data was performed for 
each system, indicated by the solid lines in Fig. \ref{fig:PBU}(a). These breakup 
functions provide a useful comparison to previous 
work. We also present an alternative parameterisation of the breakup function. A perhaps more 
intuitive way to parameterise breakup probabilities is as a function 
of the distance of closest approach relative to the average barrier radius $R_\mathrm{B}$, in the 
form of Eqn. 4 of Ref. \cite{Corradi1990}, such that 
\begin{equation}
P_\mathrm{BU} = P(R_B) e^{\mu (R_\mathrm{min}-R_\mathrm{B})},
\label{reparam}
\end{equation}
where $P(R_B)$ is the probability of breakup along a trajectory that reaches a distance of closest 
approach $R_\mathrm{B}$, and $\mu$ the 
same slope parameter as in Eqn. \ref{breakupfunction}. A detailed discussion of the 
physical significance of these parameters can be found in Ref. \cite{Corradi1990}. $R_{B}$ was 
parameterised as $R_B = 1.44(A_T^{1/3}+A_P^{1/3})$, which reproduced the $R_{B}$ of the calculated 
S\~{a}o Paulo potentials between the $^8$Be and the target-like nucleus within 0.1 fm. Where the 
target nucleus is deformed, as is the case for \er\ and \w, the breakup function is an average 
over all orientations. The resulting breakup probabilities are shown in Fig. \ref{fig:PBU}(b). From 
this, it is apparent that the dependence of breakup probability on the targets studied in this work 
is fairly small. Instead, near-target breakup is dominantly driven by how close the trajectory 
comes to $R_\mathrm{B}$. This agrees with what was found in  Ref. \cite{Rafiei2010b}. 

\begin{figure}
\begin{center}
\includegraphics[width=\columnwidth]{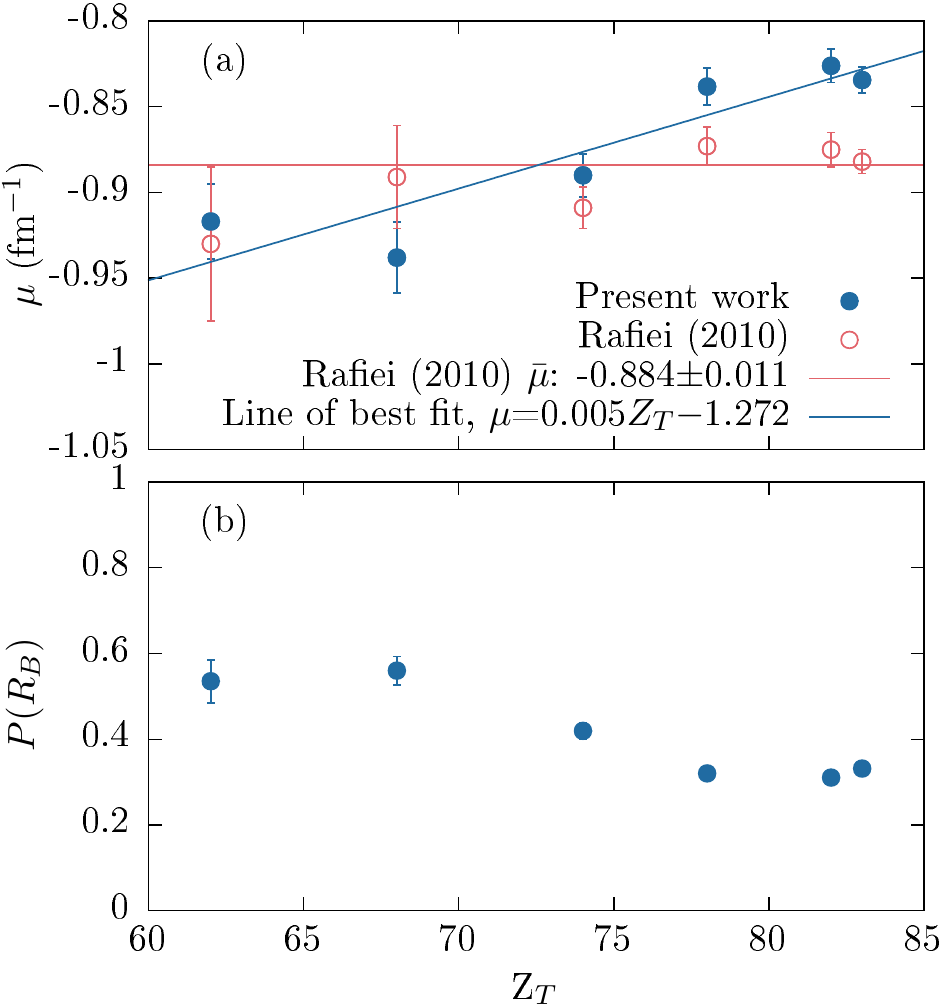}
 \caption{(Colour online) (a) Filled circles: Slope parameters $\mu \mathrm{\ (fm}^{-1})$ 
derived 
from least-squares fits to the experimental data shown in Fig. \ref{fig:PBU}(b), fit with 
$P_\mathrm{BU} = P(R_B) e^{\mu (R_\mathrm{min}-R_\mathrm{B})}$, shown as a function of target, 
$Z_T$. There is a slight $Z_T$ 
dependence on the slope, indicated by the line of best fit $\mu = 0.005 Z_T -1.272$. Open 
circles show results from Ref. \cite{Rafiei2010b}, which 
have mean slope $\bar{\mu}=-0.884 \pm 0.011$ (red line). The reasons for the discrepancies 
between the present and previous work are discussed in the text. (b) Corresponding  
$P(R_B)$  values 
derived from least-squares fits to the experimental data shown in Fig. \ref{fig:PBU}(b), using 
Eqn. \ref{reparam}. } 
 \label{fig:meanslopes}
 \end{center}
\end{figure}

The fitted breakup slope parameters using both 
parameterisations are given in Table \ref{table:params}. The reported uncertainties $\sigma$ in the 
parameters come from each least-squares fit. The parameters $\mu$ and $P(R_B)$ of the 
breakup functions are shown as a function of $Z_T$ in Fig. \ref{fig:meanslopes}. Unlike those found 
in Ref. \cite{Rafiei2010b}, there is a fairly weak $Z_T$ dependence on the fitted $\mu$ -- a 
line of best fit yields $\mu = 0.005 Z_T - 1.272$. There is also a trend of increasing $P(R_B)$ 
with decreasing $Z_T$. This is correlated with the trend of increasing ground-state neutron 
stripping Q-value with decreasing $Z_T$, as well as the number of states available for population 
near the the optimum Q-value of 0 MeV. It would be interesting to see how these trends evolve as 
$Z_T$ decreases. 

While the breakup functions derived in this work are comparable to those found by Rafiei \textit{et 
al}. 
\cite{Rafiei2010b}, there is an average increase in the probability of breakup by a factor of $1.14 
\pm 0.09$ at $R_\mathrm{min} - R_\mathrm{B}=4$ fm. These differences result from the combined 
effects 
of several factors that have been discussed above, but are summarised here: (i) the Rutherford 
scattering yield in the normalisation bin for every measurement is a factor of $0.921 
\pm 0.009$  lower due to slight refinement in the actual position of the \balin\ array, (ii) the 
 coincidence efficiency of these $\alpha-\alpha$ pairs calculated using \platypus\
with respect to $\theta_{12}$ is different to that deduced in the previous work, 
and has a different $Z_T$ and $E_\mathrm{beam}$ dependence, and (iii) correcting for 
coincidence efficiency produces an efficiency corrected yield over all 
azimuthal angles, and the calculation of the Rutherford yield must reflect this, as discussed in 
Appendix \ref{AppendixB}. As seen in Fig. \ref{fig:meanslopes}, the slope, $\mu$, of the breakup 
function becomes shallower with increasing $Z_T$. The difference in average slope from the previous 
work is 
primarily driven by the two-dimensional coincidence efficiency correction used in this work. The 
improvements to \platypus\ and to the efficiency determinations allow reliable cross-sections to 
be determined, which could be analysed using available semi-classical methods 
\cite{Knoll1977,Vigezzi1989}. With these new breakup functions, the next step is then to determine 
the impact of breakup on fusion suppression with the modified \platypus\ model. 

\begin{table}
\caption{Near-target breakup function parameters determined through least-squares fits to the 
experimental data shown in Fig. \ref{fig:PBU} for the breakup of \eightbe\ formed after neutron 
transfer in reactions of \ninebe\ with \heavy.}
\begin{center}
\begin{tabularx}{\columnwidth}{cXXXXXX}\addlinespace[3 pt]\toprule[1pt]\addlinespace[3 pt]
× & \sm & \er & \w & \pt & \pb & \bi \\\addlinespace[3 pt]
\midrule[0.5 pt]\addlinespace[3 pt]
$\mu$ (fm$^{-1}$) 	& -0.92  & -0.94 & -0.89 & -0.84 & -0.83 & -0.83\\
$\sigma_\mu$ (fm$^{-1}$)& 0.02   & 0.02  & 0.01  &  0.01 &  0.01 & 0.01 \\
$\nu$  			& 9.0    & 9.6   &9.0    &8.3    &8.3    & 8.4\\
$\sigma_\nu$ 		&0.3	 &0.3	 &0.2    &0.2    &0.2    &0.1\\
$P(R_B)$		&0.54   & 0.56 &0.42  &0.32   &0.31  &0.33\\
$\sigma_{P(R_B)}$       &0.05   &0.03  &0.02 &0.02    &0.01  &0.01\\\bottomrule[1pt]
\end{tabularx}
\end{center}
\label{table:params} 
\end{table}

\section{Above-barrier incomplete fusion cross-sections\label{FICF}}

There have been two major approaches towards characterising fusion suppression in collisions with 
 weakly bound nuclei. The first is through comparing measured above-barrier complete fusion 
cross sections to coupled-channels predictions of fusion cross-sections 
$\sigma^\mathrm{expt.}_\mathrm{CF}/\sigma^\mathrm{calc.}_\mathrm{fus.}$ (e.g. 
\cite{dasgupta1999fusion,Dasgupta2004,Gomes2006,Palshetkar2010,Parkar2010}). This approach relies on 
accurate determination of the average barrier energy \cite{Dasgupta2004} and is somewhat model 
dependant \cite{Gomes2011}. The second approach equates fusion suppression to the fraction of 
incomplete fusion to total fusion 
$F_\mathrm{ICF}=\frac{\sigma_\mathrm{ICF}}{\sigma_\mathrm{ICF}+\sigma_\mathrm{CF}}$. 
Incomplete fusion is defined experimentally as capture of only part of the charge of the projectile.
This approach is justified by measurements which find similar values for 
(1-$\sigma^\mathrm{expt.}_\mathrm{CF}/\sigma^\mathrm{calc.}_\mathrm{fus.}$) and $F_\mathrm{ICF}$ 
\cite{dasgupta1999fusion}. As such, experimental measures of $F_\mathrm{ICF}$ are thought to 
provide an indirect measure of fusion suppression that is model independent. 

When trying to understand the role of breakup in the observed suppressions 
of complete fusion, it has been conjectured that \sigmaicf\ (and thus $F_\mathrm{ICF}$) is entirely 
due to breakup of the weakly bound nucleus followed by capture of one of the fragments. However, it 
is very difficult to separate breakup followed by capture of one of the fragments from a transfer 
process forming the same nucleus. If transfer comprises a large fraction of \sigmaicf, 
$F_\mathrm{ICF}$ cannot be attributed solely to breakup. Further, \sigmaicf+\sigmacf\
can no longer be interpreted as the total fusion cross-section.  In the case of $^7$Li + 
$^{165}$Ho, exclusive measurements of $\gamma$-rays and charged fragments favour the interpretation 
that \sigmaicf\ is predominantly due to breakup \cite{Tripathi2005}. While the interpretation of 
\sigmaicf\ is ambiguous experimentally, it is clear within a classical model. By using \platypus, 
the contribution of breakup to $F_\mathrm{ICF}$ can be determined. 

\platypus\ is designed to provide predictions of \sigmacf\ and 
\sigmaicf\ at energies above the barrier, through the use of the experimentally determined breakup 
functions, applied at above-barrier energies. In \platypus, ICF is assumed to occur when one of the 
breakup fragments passes inside the barrier radius, while CF occurs when either the unbroken 
projectile or both breakup fragments pass the barrier radius. Calculations were performed using the 
near-target breakup functions determined from the least-squares fit to the below-barrier 
experimental breakup data, 
which have parameters as shown in Table \ref{table:params}. Nuclear   
potentials were calculated using the S\~{a}o Paulo potential \cite{Chamon2002}. Calculations were 
performed for partial waves up to $100\hbar$, with 
200000 breakup events simulated in total. The yield of near-target 
transfer-triggered breakup was attributed exclusively to breakup of the \twoplus\ resonance in 
\eightbe, and thus the modelled excitation energies and lifetimes of the \eightbe\ projectile were 
those of the \twoplus\ state, as shown in Fig. \ref{fig:distroEx}. Near-target breakup of \eightbe, 
in addition to arising from the \twoplus\ state, should have some contribution from the high 
excitation energy tail of the \zeroplus\ state. Test calculations show that this contribution 
should be expected to \textit{decrease} the overall $F_\mathrm{ICF}$ arising from near-target 
transfer-triggered breakup. This is because the average excitation energy of the high-energy tail 
of the \zeroplus\ state is lower than that of the \twoplus\ state, as can be seen in Fig. 
\ref{fig:distroEx}. Hence the average lifetime is  longer, and a smaller fraction of near-target 
breakup will occur prior to reaching the fusion barrier. Calculations of $F_\mathrm{ICF}$ were made 
at energies in $0.05 V_B$ steps from $1.05 - 1.30 V_B$, consistent with previous work 
\cite{Rafiei2010b}.  Over the energy 
range of $1.05 - 1.30 V_B$, $F_\mathrm{ICF}$ is energy dependent, and varies by a factor of two 
for each reaction, from F$_\mathrm{ICF}=0.16$ at $1.05 V_B$ to $0.08$ at $1.30 V_B$ on average. The 
results from each energy step have been averaged to give a $F_\mathrm{ICF}$ 
value for each system, to compare to previous work, and to experimental measures.

\begin{figure}
\begin{center}
\includegraphics[width=\columnwidth]{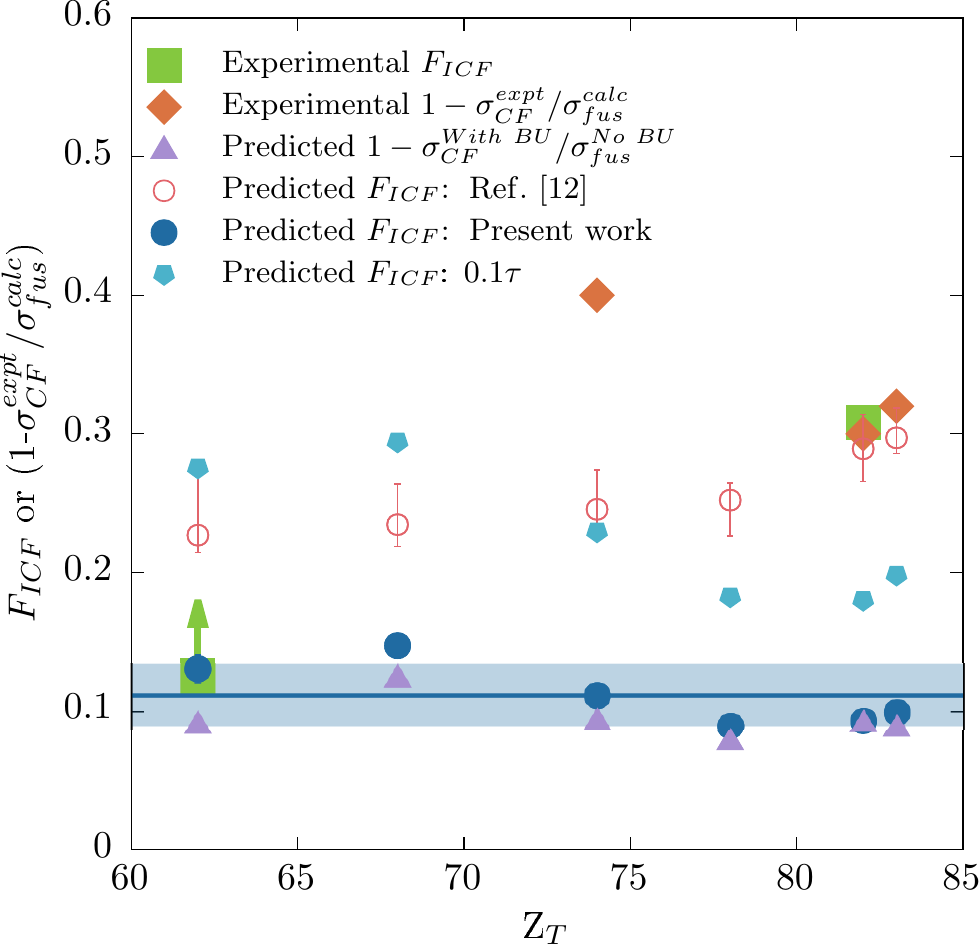}
 \caption{(Colour online) Experimental 
values of $F_\mathrm{ICF}$ \cite{Gomes2006,Dasgupta2010} (filled 
squares), and
$1-\sigma^\mathrm{expt.}_\mathrm{CF}/\sigma^\mathrm{calc.}_\mathrm{fus}$ 
\cite{Dasgupta2004,Fang2013} (filled diamonds), shown as a function of target $Z$. Predictions of 
$F_\mathrm{ICF}$ (filled circles) and complete fusion suppression (filled triangles) using the new 
breakup functions and the modified version of platypus. Error bars (determined from the uncertainty 
in the least-squares fit) are smaller than the points.  
The $F_\mathrm{ICF}$ and complete fusion suppression predictions show no clear trend with $Z$.
The $F_{ICF}$ prediction has a  mean value of $0.11 \pm 0.02$ shown as the solid line, and the 
shaded bar indicates $\pm 1 \sigma$.  $1-\sigma^\mathrm{With\ BU}_\mathrm{CF}/\sigma^\mathrm{No\ 
BU}_\mathrm{fus}$  has a mean value of $0.09 \pm 0.02$. $F_\mathrm{ICF}$  predictions made using 
the lifetime of the \twoplus\ state ten times smaller than expected are shown with pentagons. 
 $F_\mathrm{ICF}$  predictions from Ref. \cite{Rafiei2010b} are shown with open circles. } 
 \label{fig:FICF}
 \end{center}
\end{figure}

The resulting \sigmacf\ and \sigmaicf\ are presented as $F_\mathrm{ICF}$\ shown by the filled 
circles (blue) in Fig. \ref{fig:FICF}. In contrast expectations from the empirical 
prediction of Ref. \cite{Hinde2002}, these new predictions show no significant dependence on target 
$Z$ in the range studied in this work, and have a mean value of $0.11 \pm 0.02$, which is indicated 
by the solid line Fig. \ref{fig:FICF}. For comparison, the $F_\mathrm{ICF}$ predictions from Ref. 
\cite{Rafiei2010b} are shown by open circles. While several changes were made 
to the determination of coincidence efficiencies and extraction of breakup probabilities, the 
total change in the breakup functions used as input for calculations of above-barrier  
$F_\mathrm{ICF}$ was relatively modest, as already discussed. Therefore, the changes to 
\platypus\ to model breakup of \eightbe\ through the \twoplus\ resonance are the major drivers 
towards the observed reduction of $F_\mathrm{ICF}$ by a factor of 2-3 relative to Ref. 
\cite{Rafiei2010b}.


Experimentally, complete fusion suppression has been deduced, independently of \sigmaicf, through 
comparison with reactions forming the same compound nucleus involving only well bound nuclei 
\cite{Dasgupta2004,Rath2009}. Within the classical dynamical model followed in \platypus, 
$F_\mathrm{ICF}$ and complete fusion suppression are directly related, excepting some impact 
parameters outside the grazing trajectory that can only contribute to \sigmaicf\ and not to 
\sigmacf. To demonstrate that such trajectories do not make a significant contribution to 
\sigmaicf, we performed calculations with \platypus\ switching off breakup. The resulting fusion 
cross-section $\sigma_{fus}^{No\ BU}$ is compared with $\sigma_{CF}^{with\ BU}$ obtained with 
\platypus. The quantity ($1-\sigma_{CF}^{with\ BU}/\sigma_{fus}^{No\ BU}$), shown by purple 
triangles in Fig. \ref{fig:FICF}, is very close to $F_\mathrm{ICF}$. This demonstrates that 
contributions to \sigmaicf\ from trajectories outside the grazing trajectory is small. 

To understand the specific role of lifetime in $F_\mathrm{ICF}$ predictions, the lifetime of the 
\twoplus\ state was changed to be a factor of ten smaller. The results are shown by the 
blue pentagons in Fig. \ref{fig:FICF}, and are typically a factor of two larger than previously 
(blue circles). This result makes the importance of explicit handling of lifetimes very 
clear. Indeed, the experimentally measured \thetabeta\ distributions compared to \platypus\ 
simulations, shown in Fig. \ref{fig:thetabeta}, already indicates that at below-barrier energies, 
the explicit inclusion of lifetimes change the breakup observables. 

Experimental measurements of $F_\mathrm{ICF}$ (which include any contributions from transfer) 
are shown in Fig. \ref{fig:FICF} as solid squares for \ninebe\ + \pb\ \cite{Dasgupta2004} and \sm\ 
\cite{Gomes2006}. For $F_\mathrm{ICF}$ measurements to be made, both CF and ICF cross-sections must 
be measured. However, as both CF and ICF cross-sections are unavailable, fusion 
suppression factors $1-\sigma^\mathrm{expt.}_\mathrm{CF}/\sigma^\mathrm{calc.}_\mathrm{fus.}$ are 
shown for \ninebe\ + \bi\ \cite{Dasgupta2010} and \w\ 
\cite{Fang2013} as diamonds in  Fig. \ref{fig:FICF}. As both $F_\mathrm{ICF}$ and the fusion 
suppression factor are available for  \ninebe\ + \pb \cite{Dasgupta2004}, both are shown, 
demonstrating the 
agreement between both measures in this system. The measured $F_\mathrm{ICF}$ and fusion 
suppressions for \ninebe\ + \bi\ and \pb\  are a factor of three times larger than the predicted 
contribution from neutron-transfer triggered breakup, and the experimental fusion suppression 
determined for \ninebe\ + \w\ is a factor of four times larger. The $F_\mathrm{ICF}$ determined for  
\ninebe + \sm\ is consistent with the prediction. However, the measured ICF cross section in this 
experiment represents a lower limit, as cross-sections for $^{146}$Gd and $^{148}$Gd were not 
included \cite{Gomes2006}. Further, as indicated in Fig. \ref{fig:FICF}, even with lifetimes that 
are a factor of ten smaller than those estimated from the width of the \twoplus\ resonance in 
\eightbe, the predicted $F_\mathrm{ICF}$ cannot be reconciled with experiment. 


\section{Conclusions\label{conclusions}}

Explicit inclusion of excitation energies and lifetimes of unbound resonances are crucial to 
model breakup. In the absence of a quantum mechanical model of transfer-triggered breakup, they 
have been included by modifying the classical dynamical code \platypus. The new calculations show 
improved agreement  
with the measured energy and angular correlations of the breakup fragments. These correlations show 
sensitivity even to the sub-zeptosecond lifetimes of the \twoplus\ 
state of \eightbe\ formed following n-transfer from \ninebe. Above the barrier, the inclusion of 
these lifetimes significantly reduces 
predicted above-barrier suppression of complete fusion. This occurs because a larger fraction of 
nuclei remain intact until reaching the barrier. As a result, predicted complete fusion 
cross-sections are not suppressed to the extent expected from earlier calculations that do not 
explicitly include lifetimes. This result is expected to apply to weakly-bound nuclei in general. 

To make quantitative predictions of complete fusion suppression at above-barrier energies, breakup 
probabilities extracted from the experiments were used as input to the modified version of 
\platypus\ that explicitly includes lifetime effects. This results in incomplete fusion to total 
fusion fractions $F_\mathrm{ICF}$ of $\sim 11\%$ at above-barrier energies. The related complete 
fusion suppression of $\sim 9\%$ is much less than the experimentally measured $F_\mathrm{ICF}$ and 
complete fusion suppressions  of $30-40\%$. \cite{Fang2013,Dasgupta2004,Dasgupta2010} 

Three key conclusions are drawn from these results:

(1) As the calculated $F_\mathrm{ICF}$ is much less than measured, the cross-sections that are 
attributed experimentally to ICF may include a significant 
contribution from transfer directly producing the same heavy nucleus. This needs to be investigated 
in more detail. 

(2) If \sigmaicf\ contains contributions from both ICF and 
transfer, defining an empirical complete fusion suppression $F_{ICF}$ in 
terms of \sigmaicf\ is problematic. 

(3) The observed reduction of complete fusion at above-barrier energies has been measured 
independently of \sigmaicf\ in several reactions through direct comparison  with reactions of well 
bound nuclei \cite{Dasgupta2004, Rath2009}. Since breakup cannot explain this, then other processes 
must contribute. Experimental values of $F_\mathrm{ICF}$ and 
$1-\sigma^\mathrm{expt.}_\mathrm{CF}/\sigma^\mathrm{calc.}_\mathrm{fus.}$ have 
been found to be similar \cite{Dasgupta2004}, thus it is reasonable to suspect that the two 
quantities are linked. Therefore, if transfer is shown to make a large contribution to products 
previously attributed to ICF, then a mechanism by which transfer may suppress complete fusion needs 
to be considered. In a classical picture, if transfer removes energy from the relative motion, it 
will reduce fusion. However in a coupled-channels approach, it is not clear whether above-barrier 
fusion can be suppressed by transfer. These questions require further investigation. 

\begin{acknowledgments}
This work was supported by ARC grants FL110100098, DP130101569 and DP140101337. Support for 
ANU Heavy Ion Accelerator 
Facility accelerator operations through the NCRIS program is acknowledged. Helpful discussions with 
A. Diaz-Torres and R. Rafiei 
are gratefully acknowledged. We thank R. du Rietz and M. Evers for their assistance in running the 
original experiment.\end{acknowledgments}

\appendix

\section{Removal of cross-talk events\label{AppendixA}}

In the previous work, spurious coincident events resulting from 
charge-sharing 
across adjacent pixels resulting from cross-talk or particles crossing the inter-strip 
partition, were 
removed by rejecting \textit{any} event in adjacent pixels. With greater experience in analysis of 
such data, it was realised that these events can be 
rejected by their unphysical relative energy (\erel), with respect to their opening angle 
(\thetaonetwo). In this analysis, spurious events were removed by applying cuts in the \erel -- 
\thetaonetwo\ spectra. This alternate method for extracting breakup events resulted in an 
approximately four times larger yield of the ground-state \eightbe\ events, as the vast majority of 
genuine \eightbe\
ground-state breakup events result in signals in adjacent pixels. These lost events would otherwise 
have had to be restored by a larger efficiency correction.

\section{Normalisation\label{AppendixB}}
The expected yield of Rutherford scattering $N_\mathrm{Ruth}(\theta_\mathrm{Ruth})$ may be 
determined from the yield of Rutherford scattered particles  
$N_\textrm{Ruth}(\theta_\mathrm{norm})$ in the $ \theta_\mathrm{norm} = 124^\circ$ to $127^\circ$ 
elastic normalisation bin,
 \begin{equation}
 N_\textrm{Ruth}(\theta_\mathrm{Ruth}) = 
N_\textrm{Ruth}(\theta_\mathrm{norm})\frac{(\frac{d\sigma}{d\Omega})(\theta_\mathrm{bin})}{(\frac{
d\sigma } { d\Omega }
 )(\theta_\textrm{norm})}(\frac{d\Omega_\textrm{bin}}{d\Omega_\textrm{norm}}),
 \end{equation}
where $\frac{d\sigma}{d\Omega}(\theta_{x})$ and $d\Omega(\theta_{x})$ are the 
differential cross-sections and solid angles respectively. As the efficiency corrected breakup yield 
corresponds to the number of coincidence breakup events over all azimuthal angles, the calculated 
Rutherford yield must be for this same angular range. In Ref. \cite{Rafiei2010b} the Rutherford 
yield was calculated within the coverage of \balin. This leads to a downwards correction in 
the present study by a factor of $\sim 0.75$ equal to the fractional coverage of the \balin\ array 
in azimuthal angle. 

As $\theta_\mathrm{norm}$ is at a relatively backwards angle, the elastic yield is purely 
Rutherford only for deep sub-barrier measurements. Where measurements were made near to the 
barrier, the expected $N_\textrm{Ruth}(\theta_\textrm{norm})$ was calculated 
from the elastic yield, $ N_\textrm{elas}(\theta_\textrm{norm})$, by taking the ratio of the 
elastic and Rutherford cross sections determined from optical model fits of existing elastic 
scattering data \cite{Signorini2002, Parkar2013, Yu2010, Zagrebaev}, such that 
 \begin{equation}
N_\textrm{Ruth}(\theta_\textrm{norm})= N_\textrm{elas}(\theta_\textrm{norm})
(\frac{\frac{d\sigma_{elastic}}{d\Omega}}{\frac{d\sigma_{Rutherford}}{d\Omega}})(\theta_\textrm{norm
}).
\end{equation}
The correction was largest for \ninebe\ + \pb\ and \bi\ at $E_{beam}=37$ MeV, where 
$\frac{d\sigma_{elastic}/d\Omega}{d\sigma_{Rutherford}/d\Omega}(\theta_\textrm{norm
})=0.89$. The solid angle coverage of the normalisation bin $d\Omega_\textrm{norm}$ can be 
determined from  the solid angle coverage of BALiN by comparing the yields in normalisation bin and 
in each $\theta$ bin of BALiN at a beam energy $E_{cal}$ where the elastic yields
 do not significantly deviate from Rutherford scattering for all angles. In that case, we can write 
  \begin{equation}
 d\Omega_\textrm{norm}  =
\frac{N_\textrm{norm}(\theta_\textrm{norm},E_\textrm{cal})}{N_\textrm{Ruth}(\theta_\textrm{bin},
E_\textrm {cal})}\frac{\frac{d\sigma 
}{d\Omega}_\textrm{Ruth}(\theta_\textrm{bin},E_\textrm{cal})}{\frac{d\sigma}{d\Omega}_\textrm{Ruth}
(\theta_\textrm{norm},E_\textrm{cal})} d\Omega_\textrm{bin} \label{eqn:omegaM},
 \end{equation}
where $d\Omega_\textrm{bin}$ is the solid angle coverage for each $\theta$ bin in 
\balin. 

\section{Efficiency Determination \label{Appendix:Efficiencies}}
\begin{figure*}
\includegraphics[width=\textwidth]{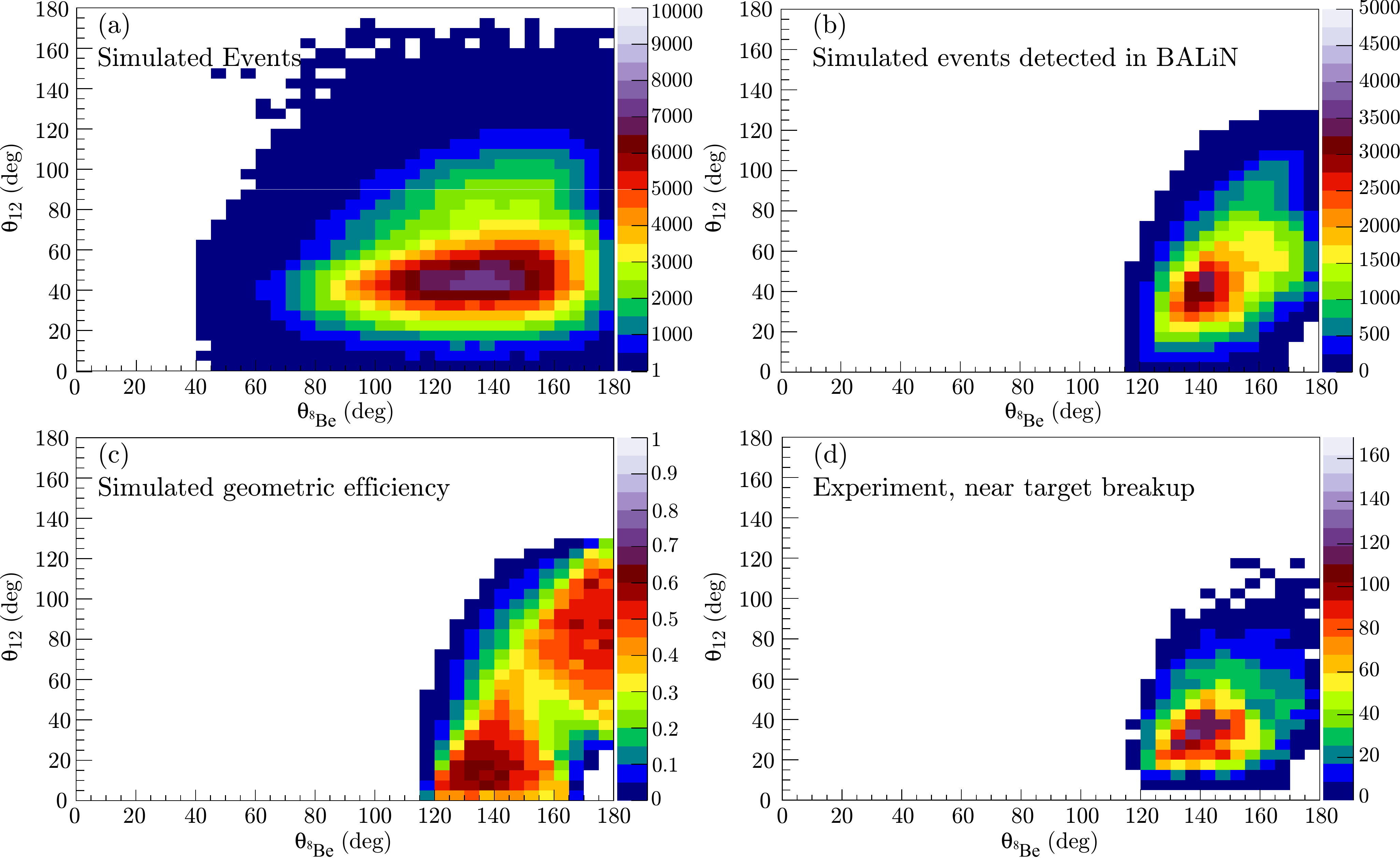}
\caption{(Colour online) Simulated and experimental near-target 
$\theta_{^8\mathrm{Be}}-\theta_\mathrm{12}$ 
distributions for $^8$Be$_{2^+}$ + \pb\ $\rightarrow  \alpha + \alpha + ^{207}$Pb at $E_{Beam} =$34 
MeV. (a) Total simulated distribution. (b) The same events 
filtered by the acceptance of the \balin\ array. (c) The associated geometric 
coincidence efficiency of the \balin\ array determined from the simulated events filtered by the 
acceptance of \balin\ divided by the total simulated events in each 
($\theta_{^8\mathrm{Be}},\theta_\mathrm{12}$) bin. (d) Experimental 
$\theta_{^8\mathrm{Be}}-\theta_{12}$ 
distribution for near-target breakup events [region \textit{(ii)} of Fig. \ref{fig:QErel}] showing 
the good correspondence between the filtered simulated data and the experiment}
\label{fig:effeg}  
\end{figure*}

Using the notion of $\theta_{^8\mathrm{Be}}$, the geometric coincidence efficiency 
was given by the ratio of simulated breakup events that would have landed in \balin\ at each 
$\theta_{^8\mathrm{Be}}$ and \thetaonetwo, (taking into account the azimuthal coverage of \balin)
to the simulated events distributed over all azimuthal angles. The simulated events were subject to 
the same detector conditions as the experimental data.  
As an example, the geometric coincidence 
efficiency matrix determined for \ninebe\ + \pb\ at 34.0 MeV is shown in Fig. \ref{fig:effeg}(c). 
This was determined from the ratio of the number of events within the acceptance of \balin,  Fig. 
\ref{fig:effeg}(b), to the total number of events, Fig. \ref{fig:effeg}(a), in each 
($\theta_{^8\mathrm{Be}}$, $\theta_{12}$) bin. The experimentally determined 
($\theta_{^8\mathrm{Be}}$, $\theta_{12}$) distribution for the same system is shown in Fig. 
\ref{fig:effeg}(d). The geometric coincidence efficiency shows two triangular regions of high 
detector efficiency -- at small $\theta_{12} \sim 10$\textdegree\ with $\theta_{^8\mathrm{Be}}\sim 
135$\textdegree\ corresponding to the centre of the \balin\ array, and at $\theta_{12} \sim 
80$\textdegree\ at backward $\theta_{^8\mathrm{Be}}\sim 180$\textdegree. The former is due to 
events with sufficiently small opening angle so that both fragments land on the same DSSD, while 
the latter is due to events that strike two different DSSDs. For values of $\theta_{^8\mathrm{Be}}$ 
where \balin\ gives coverage, for some values of $\theta_{12}$ the efficiency is zero. A correction 
to account for this is made in the next stage in the determination of the efficiency. 

To simulate the distribution of fragments, shown Fig. \ref{fig:effeg}(a), 
needed for this second part of the efficiency correction, \platypus\ calculations using the 
modifications discussed in Sec. \ref{platypus} were performed. Simulations of near-target 
(high \erel) breakup events were performed using the excitation energy 
and excitation energy dependent mean life for \eightbe\ \twoplus\ as discussed in Sec. 
\ref{platypus}.
\qval\ distributions were taken from the experimental results, and the energy of the \eightbe\
pseudo-projectile ($E'_\mathrm{P}$) calculated by matching the distance of closest approach to that 
attained by the \ninebe\ beam with energy $E_\mathrm{P}$, as \platypus\ does not simulate transfer. 
In analogy to the optimum Q-value of \cite{Schiffer1973}, this matching energy is given by
\begin{equation}
E'_\mathrm{P} = E_\mathrm{P} \frac{m_\mathrm{T}}{m'_\mathrm{T}} 
\left(\frac{Z_\mathrm{P} Z_\mathrm{T}}{Z'_\mathrm{P} Z'_\mathrm{T}} 
\right),
\end{equation}
where $Z_\mathrm{T},m_\mathrm{T},Z_\mathrm{P},m_\mathrm{P}$ and 
$Z'_\mathrm{T},m'_\mathrm{T},Z'_\mathrm{P},m'_\mathrm{P}$ is the charge and 
mass of the target and projectile before and after transfer, respectively. In this case, where only 
neutron transfer is occurring, the matching energy is very close to the experimental beam energy. 
In 
cases such as the breakup of $^7$Li, where proton transfer dominates, this factor becomes more 
important. The projectile-target and fragment-target potentials 
are Woods-Saxon parameterisations of S\~{a}o Paulo potentials \cite{Chamon2002}, from 
\cite{Rafiei2010b}.  This is used for all \platypus\ 
simulations in this work. According to these simulations, events that have $\theta_{^8\mathrm{Be}}$ 
where \balin\ gives coverage, but have $\theta_{12}$ where the efficiency is zero accounted 
for $\sim 7 \%$ of all 
events simulated within the $\theta_{^8\mathrm{Be}}$ acceptance of the array. As such, this second 
step in efficiency correction represents a small (though $\theta_{^8\mathrm{Be}}$ dependant) 
model-dependent addition to a
model-independent efficiency correction.

\section{Comparison of efficiencies in Ref. \cite{Rafiei2010b} and the present work 
\label{Appendix:Comparison}}
The coincidence efficiencies determined in this work differ from those found in Ref. 
\cite{Rafiei2010b}: there, prompt efficiencies were 
given for two $15^\circ$ bins in  $\theta_{^8\mathrm{Be}}$, $\epsilon_{121-136^\circ}=0.25$ and 
$\epsilon_{136-151^\circ}=0.42$, 
and the efficiencies were found to be nearly independent of $E_{lab}$. Here, the efficiencies are 
calculated in $5^\circ$ bins in $(\theta_{12},\theta_{^8\mathrm{Be}})$, and so are much more 
fine-grained. However, when averaged over the same range of $\theta_{^8\mathrm{Be}}$, the 
efficiencies in this work are on average $\epsilon_{121-136^\circ}=0.17$ 
and $\epsilon_{136-151^\circ}=0.40$. As the efficiency corrections in this work take into account 
$\theta_{12}$, the distribution of which changes with $E_{lab}$ and $Z$, these averaged 
efficiencies are not independent of $E_{lab}$ or $Z$. 

These differences between Ref. \cite{Rafiei2010b} and the present work can be accounted by three 
factors: (i) In the previous analysis, 
efficiencies were calculated 
as a function only of $\theta_{^8\mathrm{Be}}$. As seen in Fig. \ref{fig:effeg}(c), for events 
with a $(\theta_{12},\theta_{^8\mathrm{Be}})$ distribution as shown in Fig. \ref{fig:effeg}(a), the 
efficiency varies strongly as 
a function of $\theta_{12}$ for fixed $\theta_{^8\mathrm{Be}}$. Thus, efficiency correction only as 
a function of $\theta_{^8\mathrm{Be}}$ results in an average over-correction in the number of 
breakup pairs for each $\theta_{^8\mathrm{Be}}$ by a factor of $\sim 1.1$ for the systems studied 
in this work (depending on $\theta_{^8\mathrm{Be}}$, target mass and beam energy), compared to the 
new two-dimensional efficiency correction performed here. (ii) The 
efficiencies further change as the early version of \platypus\ used in Ref. \cite{Rafiei2010b} did 
not have a fully isotropic distribution of initial fragment directions: 
there was an over-abundance of events with similar scattering angles, $\theta_1 \sim \theta_2$, 
leading to an artificially high efficiency. \platypus\ was corrected in late 2010 
\cite{Diaz-Torrespc}. (iii) The modifications of \platypus\ performed for 
this work resulted in a different angular distribution of fragments and so changed the 
model-dependent stage of the efficiency corrections.

\section{Local Breakup Probabilities \label{appendixPBU}} 
Since interacting nuclei spend more time near the distance of closest approach, then
casting the breakup probability as 
\begin{equation}
\frac{dP}{dr} \propto e^{\mu r},
\end{equation}
in a dynamical model is inappropriate. To illustrate this, consider a classical Coulomb trajectory, 
where
\begin{equation}
\frac{dt}{dr} = \frac{r}{v \sqrt{(r-a_0(1+\epsilon))(r-a_0(1-\epsilon))}}, 
\end{equation}
\noindent and $a_0 = Z_p Z_t e^2/\mu v^2$, $\epsilon = \sqrt{1+(L/\eta)^2}$ and the 
Sommerfeld parameter $\eta = Z_p Z_t 
e^2 /v$, where $\mu$ is the reduced mass, and $v$ the incident velocity. Then,
\begin{equation}
\frac{dP}{dt} = \frac{dP}{dr}\frac{dr}{dt} \propto e^{-\mu r} 
\frac{v \sqrt{(r-a_0(1+\epsilon))(r-a_0(1-\epsilon))}}{r}.
\end{equation}
\noindent For a trajectory corresponding to scattering at $180^\circ$, $\epsilon = 1$ and the 
distance of closest approach, $R_0 = 
2a_0$, this results in $dP(R_{min})/dt=0$ at the distance of closest approach, which does not seem 
reasonable.

\bibliography{HeavyPaper.bib}

\begin{thebibliography}{35}%
\makeatletter
\providecommand \@ifxundefined [1]{%
 \@ifx{#1\undefined}
}%
\providecommand \@ifnum [1]{%
 \ifnum #1\expandafter \@firstoftwo
 \else \expandafter \@secondoftwo
 \fi
}%
\providecommand \@ifx [1]{%
 \ifx #1\expandafter \@firstoftwo
 \else \expandafter \@secondoftwo
 \fi
}%
\providecommand \natexlab [1]{#1}%
\providecommand \enquote  [1]{``#1''}%
\providecommand \bibnamefont  [1]{#1}%
\providecommand \bibfnamefont [1]{#1}%
\providecommand \citenamefont [1]{#1}%
\providecommand \href@noop [0]{\@secondoftwo}%
\providecommand \href [0]{\begingroup \@sanitize@url \@href}%
\providecommand \@href[1]{\@@startlink{#1}\@@href}%
\providecommand \@@href[1]{\endgroup#1\@@endlink}%
\providecommand \@sanitize@url [0]{\catcode `\\12\catcode `\$12\catcode
  `\&12\catcode `\#12\catcode `\^12\catcode `\_12\catcode `\%12\relax}%
\providecommand \@@startlink[1]{}%
\providecommand \@@endlink[0]{}%
\providecommand \url  [0]{\begingroup\@sanitize@url \@url }%
\providecommand \@url [1]{\endgroup\@href {#1}{\urlprefix }}%
\providecommand \urlprefix  [0]{URL }%
\providecommand \Eprint [0]{\href }%
\providecommand \doibase [0]{http://dx.doi.org/}%
\providecommand \selectlanguage [0]{\@gobble}%
\providecommand \bibinfo  [0]{\@secondoftwo}%
\providecommand \bibfield  [0]{\@secondoftwo}%
\providecommand \translation [1]{[#1]}%
\providecommand \BibitemOpen [0]{}%
\providecommand \bibitemStop [0]{}%
\providecommand \bibitemNoStop [0]{.\EOS\space}%
\providecommand \EOS [0]{\spacefactor3000\relax}%
\providecommand \BibitemShut  [1]{\csname bibitem#1\endcsname}%
\let\auto@bib@innerbib\@empty
\bibitem [{\citenamefont {Dasgupta}\ \emph {et~al.}(1999)\citenamefont
  {Dasgupta}, \citenamefont {Hinde}, \citenamefont {Butt}, \citenamefont
  {Anjos}, \citenamefont {Berriman}, \citenamefont {Carlin}, \citenamefont
  {Gomes}, \citenamefont {Morton}, \citenamefont {Newton}, \citenamefont
  {{Szanto de Toledo}},\ and\ \citenamefont {Hagino}}]{dasgupta1999fusion}%
  \BibitemOpen
  \bibfield  {author} {\bibinfo {author} {\bibfnamefont {M.}~\bibnamefont
  {Dasgupta}}, \bibinfo {author} {\bibfnamefont {D.~J.}\ \bibnamefont {Hinde}},
  \bibinfo {author} {\bibfnamefont {R.~D.}\ \bibnamefont {Butt}}, \bibinfo
  {author} {\bibfnamefont {R.~M.}\ \bibnamefont {Anjos}}, \bibinfo {author}
  {\bibfnamefont {A.~C.}\ \bibnamefont {Berriman}}, \bibinfo {author}
  {\bibfnamefont {N.}~\bibnamefont {Carlin}}, \bibinfo {author} {\bibfnamefont
  {P.~R.~S.}\ \bibnamefont {Gomes}}, \bibinfo {author} {\bibfnamefont {C.~R.}\
  \bibnamefont {Morton}}, \bibinfo {author} {\bibfnamefont {J.~O.}\
  \bibnamefont {Newton}}, \bibinfo {author} {\bibfnamefont {A.}~\bibnamefont
  {{Szanto de Toledo}}}, \ and\ \bibinfo {author} {\bibfnamefont
  {K.}~\bibnamefont {Hagino}},\ }\href {\doibase 10.1103/PhysRevLett.82.1395}
  {\bibfield  {journal} {\bibinfo  {journal} {Physical Review Letters}\
  }\textbf {\bibinfo {volume} {82}},\ \bibinfo {pages} {1395} (\bibinfo {year}
  {1999})}\BibitemShut {NoStop}%
\bibitem [{\citenamefont {Dasgupta}\ \emph {et~al.}(2004)\citenamefont
  {Dasgupta}, \citenamefont {Gomes}, \citenamefont {Hinde}, \citenamefont
  {Moraes}, \citenamefont {Anjos}, \citenamefont {Berriman}, \citenamefont
  {Butt}, \citenamefont {Carlin}, \citenamefont {Lubian}, \citenamefont
  {Morton}, \citenamefont {Newton},\ and\ \citenamefont {{Szanto de
  Toledo}}}]{Dasgupta2004}%
  \BibitemOpen
  \bibfield  {author} {\bibinfo {author} {\bibfnamefont {M.}~\bibnamefont
  {Dasgupta}}, \bibinfo {author} {\bibfnamefont {P.~R.~S.}\ \bibnamefont
  {Gomes}}, \bibinfo {author} {\bibfnamefont {D.~J.}\ \bibnamefont {Hinde}},
  \bibinfo {author} {\bibfnamefont {S.~B.}\ \bibnamefont {Moraes}}, \bibinfo
  {author} {\bibfnamefont {R.~M.}\ \bibnamefont {Anjos}}, \bibinfo {author}
  {\bibfnamefont {A.~C.}\ \bibnamefont {Berriman}}, \bibinfo {author}
  {\bibfnamefont {R.~D.}\ \bibnamefont {Butt}}, \bibinfo {author}
  {\bibfnamefont {N.}~\bibnamefont {Carlin}}, \bibinfo {author} {\bibfnamefont
  {J.}~\bibnamefont {Lubian}}, \bibinfo {author} {\bibfnamefont {C.~R.}\
  \bibnamefont {Morton}}, \bibinfo {author} {\bibfnamefont {J.~O.}\
  \bibnamefont {Newton}}, \ and\ \bibinfo {author} {\bibfnamefont
  {A.}~\bibnamefont {{Szanto de Toledo}}},\ }\href
  {http://prc.aps.org/abstract/PRC/v70/i2/e024606} {\bibfield  {journal}
  {\bibinfo  {journal} {Physical Review C}\ }\textbf {\bibinfo {volume} {70}},\
  \bibinfo {pages} {024606} (\bibinfo {year} {2004})}\BibitemShut {NoStop}%
\bibitem [{\citenamefont {Signorini}\ \emph {et~al.}(2004)\citenamefont
  {Signorini}, \citenamefont {Glodariu}, \citenamefont {Liu}, \citenamefont
  {Mazzocco}, \citenamefont {Ruan},\ and\ \citenamefont
  {Soramel}}]{Signorini2004}%
  \BibitemOpen
  \bibfield  {author} {\bibinfo {author} {\bibfnamefont {C.}~\bibnamefont
  {Signorini}}, \bibinfo {author} {\bibfnamefont {T.}~\bibnamefont {Glodariu}},
  \bibinfo {author} {\bibfnamefont {Z.~H.}\ \bibnamefont {Liu}}, \bibinfo
  {author} {\bibfnamefont {M.}~\bibnamefont {Mazzocco}}, \bibinfo {author}
  {\bibfnamefont {M.}~\bibnamefont {Ruan}}, \ and\ \bibinfo {author}
  {\bibfnamefont {F.}~\bibnamefont {Soramel}},\ }\href@noop {} {\bibfield
  {journal} {\bibinfo  {journal} {Progress of Theoretical Physics Supplement}\
  }\textbf {\bibinfo {volume} {154}},\ \bibinfo {pages} {272} (\bibinfo {year}
  {2004})}\BibitemShut {NoStop}%
\bibitem [{\citenamefont {Dasgupta}\ \emph {et~al.}(2010)\citenamefont
  {Dasgupta}, \citenamefont {Hinde}, \citenamefont {Sheehy},\ and\
  \citenamefont {Bouriquet}}]{Dasgupta2010}%
  \BibitemOpen
  \bibfield  {author} {\bibinfo {author} {\bibfnamefont {M.}~\bibnamefont
  {Dasgupta}}, \bibinfo {author} {\bibfnamefont {D.~J.}\ \bibnamefont {Hinde}},
  \bibinfo {author} {\bibfnamefont {S.~L.}\ \bibnamefont {Sheehy}}, \ and\
  \bibinfo {author} {\bibfnamefont {B.}~\bibnamefont {Bouriquet}},\ }\href
  {\doibase 10.1103/PhysRevC.81.024608} {\bibfield  {journal} {\bibinfo
  {journal} {Physical Review C}\ }\textbf {\bibinfo {volume} {81}},\ \bibinfo
  {pages} {024608} (\bibinfo {year} {2010})}\BibitemShut {NoStop}%
\bibitem [{\citenamefont {Dasgupta}\ \emph {et~al.}(2002)\citenamefont
  {Dasgupta}, \citenamefont {Hinde}, \citenamefont {Hagino}, \citenamefont
  {Moraes}, \citenamefont {Gomes}, \citenamefont {Anjos}, \citenamefont {Butt},
  \citenamefont {Berriman}, \citenamefont {Carlin}, \citenamefont {Morton},
  \citenamefont {Newton},\ and\ \citenamefont {{Szanto de
  Toledo}}}]{Dasgupta2002}%
  \BibitemOpen
  \bibfield  {author} {\bibinfo {author} {\bibfnamefont {M.}~\bibnamefont
  {Dasgupta}}, \bibinfo {author} {\bibfnamefont {D.~J.}\ \bibnamefont {Hinde}},
  \bibinfo {author} {\bibfnamefont {K.}~\bibnamefont {Hagino}}, \bibinfo
  {author} {\bibfnamefont {S.~B.}\ \bibnamefont {Moraes}}, \bibinfo {author}
  {\bibfnamefont {P.~R.~S.}\ \bibnamefont {Gomes}}, \bibinfo {author}
  {\bibfnamefont {R.~M.}\ \bibnamefont {Anjos}}, \bibinfo {author}
  {\bibfnamefont {R.~D.}\ \bibnamefont {Butt}}, \bibinfo {author}
  {\bibfnamefont {A.~C.}\ \bibnamefont {Berriman}}, \bibinfo {author}
  {\bibfnamefont {N.}~\bibnamefont {Carlin}}, \bibinfo {author} {\bibfnamefont
  {C.~R.}\ \bibnamefont {Morton}}, \bibinfo {author} {\bibfnamefont {J.~O.}\
  \bibnamefont {Newton}}, \ and\ \bibinfo {author} {\bibfnamefont
  {A.}~\bibnamefont {{Szanto de Toledo}}},\ }\href {\doibase
  10.1103/PhysRevC.66.041602} {\bibfield  {journal} {\bibinfo  {journal}
  {Physical Review C}\ }\textbf {\bibinfo {volume} {66}},\ \bibinfo {pages}
  {041602(R)} (\bibinfo {year} {2002})}\BibitemShut {NoStop}%
\bibitem [{\citenamefont {Rath}\ \emph {et~al.}(2009)\citenamefont {Rath},
  \citenamefont {Santra}, \citenamefont {Singh}, \citenamefont {Tripathi},
  \citenamefont {Parkar}, \citenamefont {Nayak}, \citenamefont {Mahata},
  \citenamefont {Palit}, \citenamefont {Kumar}, \citenamefont {Mukherjee},
  \citenamefont {Appannababu},\ and\ \citenamefont {Choudhury}}]{Rath2009}%
  \BibitemOpen
  \bibfield  {author} {\bibinfo {author} {\bibfnamefont {P.~K.}\ \bibnamefont
  {Rath}}, \bibinfo {author} {\bibfnamefont {S.}~\bibnamefont {Santra}},
  \bibinfo {author} {\bibfnamefont {N.~L.}\ \bibnamefont {Singh}}, \bibinfo
  {author} {\bibfnamefont {R.}~\bibnamefont {Tripathi}}, \bibinfo {author}
  {\bibfnamefont {V.~V.}\ \bibnamefont {Parkar}}, \bibinfo {author}
  {\bibfnamefont {B.~K.}\ \bibnamefont {Nayak}}, \bibinfo {author}
  {\bibfnamefont {K.}~\bibnamefont {Mahata}}, \bibinfo {author} {\bibfnamefont
  {R.}~\bibnamefont {Palit}}, \bibinfo {author} {\bibfnamefont
  {S.}~\bibnamefont {Kumar}}, \bibinfo {author} {\bibfnamefont
  {S.}~\bibnamefont {Mukherjee}}, \bibinfo {author} {\bibfnamefont
  {S.}~\bibnamefont {Appannababu}}, \ and\ \bibinfo {author} {\bibfnamefont
  {R.~K.}\ \bibnamefont {Choudhury}},\ }\href {\doibase
  10.1103/PhysRevC.79.051601} {\bibfield  {journal} {\bibinfo  {journal}
  {Physical Review C}\ }\textbf {\bibinfo {volume} {79}},\ \bibinfo {pages}
  {051601} (\bibinfo {year} {2009})}\BibitemShut {NoStop}%
\bibitem [{\citenamefont {Gomes}\ \emph {et~al.}(2006)\citenamefont {Gomes},
  \citenamefont {Padron}, \citenamefont {Crema}, \citenamefont {Capurro},
  \citenamefont {{Fern{\'{a}}ndez Niello}}, \citenamefont {Arazi},
  \citenamefont {Mart{\'{\i}}}, \citenamefont {Lubian}, \citenamefont {Trotta},
  \citenamefont {Pacheco}, \citenamefont {Testoni}, \citenamefont
  {Rodr{\'{\i}}guez}, \citenamefont {Ortega}, \citenamefont {Chamon},
  \citenamefont {Anjos}, \citenamefont {Veiga}, \citenamefont {Dasgupta},
  \citenamefont {Hinde},\ and\ \citenamefont {Hagino}}]{Gomes2006}%
  \BibitemOpen
  \bibfield  {author} {\bibinfo {author} {\bibfnamefont {P.~R.~S.}\
  \bibnamefont {Gomes}}, \bibinfo {author} {\bibfnamefont {I.}~\bibnamefont
  {Padron}}, \bibinfo {author} {\bibfnamefont {E.}~\bibnamefont {Crema}},
  \bibinfo {author} {\bibfnamefont {O.~A.}\ \bibnamefont {Capurro}}, \bibinfo
  {author} {\bibfnamefont {J.~O.}\ \bibnamefont {{Fern{\'{a}}ndez Niello}}},
  \bibinfo {author} {\bibfnamefont {A.}~\bibnamefont {Arazi}}, \bibinfo
  {author} {\bibfnamefont {G.~V.}\ \bibnamefont {Mart{\'{\i}}}}, \bibinfo
  {author} {\bibfnamefont {J.}~\bibnamefont {Lubian}}, \bibinfo {author}
  {\bibfnamefont {M.}~\bibnamefont {Trotta}}, \bibinfo {author} {\bibfnamefont
  {A.~J.}\ \bibnamefont {Pacheco}}, \bibinfo {author} {\bibfnamefont {J.~E.}\
  \bibnamefont {Testoni}}, \bibinfo {author} {\bibfnamefont {M.~D.}\
  \bibnamefont {Rodr{\'{\i}}guez}}, \bibinfo {author} {\bibfnamefont {M.~E.}\
  \bibnamefont {Ortega}}, \bibinfo {author} {\bibfnamefont {L.~C.}\
  \bibnamefont {Chamon}}, \bibinfo {author} {\bibfnamefont {R.~M.}\
  \bibnamefont {Anjos}}, \bibinfo {author} {\bibfnamefont {R.}~\bibnamefont
  {Veiga}}, \bibinfo {author} {\bibfnamefont {M.}~\bibnamefont {Dasgupta}},
  \bibinfo {author} {\bibfnamefont {D.~J.}\ \bibnamefont {Hinde}}, \ and\
  \bibinfo {author} {\bibfnamefont {K.}~\bibnamefont {Hagino}},\ }\href
  {\doibase 10.1103/PhysRevC.73.064606} {\bibfield  {journal} {\bibinfo
  {journal} {Physical Review C}\ }\textbf {\bibinfo {volume} {73}},\ \bibinfo
  {pages} {064606} (\bibinfo {year} {2006})}\BibitemShut {NoStop}%
\bibitem [{\citenamefont {Palshetkar}\ \emph {et~al.}(2010)\citenamefont
  {Palshetkar}, \citenamefont {Santra}, \citenamefont {Chatterjee},
  \citenamefont {Ramachandran}, \citenamefont {Thakur}, \citenamefont {Pandit},
  \citenamefont {Mahata}, \citenamefont {Shrivastava}, \citenamefont {Parkar},\
  and\ \citenamefont {Nanal}}]{Palshetkar2010}%
  \BibitemOpen
  \bibfield  {author} {\bibinfo {author} {\bibfnamefont {C.~S.}\ \bibnamefont
  {Palshetkar}}, \bibinfo {author} {\bibfnamefont {S.}~\bibnamefont {Santra}},
  \bibinfo {author} {\bibfnamefont {A.}~\bibnamefont {Chatterjee}}, \bibinfo
  {author} {\bibfnamefont {K.}~\bibnamefont {Ramachandran}}, \bibinfo {author}
  {\bibfnamefont {S.}~\bibnamefont {Thakur}}, \bibinfo {author} {\bibfnamefont
  {S.~K.}\ \bibnamefont {Pandit}}, \bibinfo {author} {\bibfnamefont
  {K.}~\bibnamefont {Mahata}}, \bibinfo {author} {\bibfnamefont
  {A.}~\bibnamefont {Shrivastava}}, \bibinfo {author} {\bibfnamefont {V.~V.}\
  \bibnamefont {Parkar}}, \ and\ \bibinfo {author} {\bibfnamefont
  {V.}~\bibnamefont {Nanal}},\ }\href {\doibase 10.1103/PhysRevC.82.044608}
  {\bibfield  {journal} {\bibinfo  {journal} {Physical Review C}\ }\textbf
  {\bibinfo {volume} {82}},\ \bibinfo {pages} {044608} (\bibinfo {year}
  {2010})}\BibitemShut {NoStop}%
\bibitem [{\citenamefont {Parkar}\ \emph {et~al.}(2010)\citenamefont {Parkar},
  \citenamefont {Palit}, \citenamefont {Sharma}, \citenamefont {Naidu},
  \citenamefont {Santra}, \citenamefont {Joshi}, \citenamefont {Rath},
  \citenamefont {Mahata}, \citenamefont {Ramachandran}, \citenamefont
  {Trivedi},\ and\ \citenamefont {Raghav}}]{Parkar2010}%
  \BibitemOpen
  \bibfield  {author} {\bibinfo {author} {\bibfnamefont {V.~V.}\ \bibnamefont
  {Parkar}}, \bibinfo {author} {\bibfnamefont {R.}~\bibnamefont {Palit}},
  \bibinfo {author} {\bibfnamefont {S.~K.}\ \bibnamefont {Sharma}}, \bibinfo
  {author} {\bibfnamefont {B.~S.}\ \bibnamefont {Naidu}}, \bibinfo {author}
  {\bibfnamefont {S.}~\bibnamefont {Santra}}, \bibinfo {author} {\bibfnamefont
  {P.~K.}\ \bibnamefont {Joshi}}, \bibinfo {author} {\bibfnamefont {P.~K.}\
  \bibnamefont {Rath}}, \bibinfo {author} {\bibfnamefont {K.}~\bibnamefont
  {Mahata}}, \bibinfo {author} {\bibfnamefont {K.}~\bibnamefont
  {Ramachandran}}, \bibinfo {author} {\bibfnamefont {T.}~\bibnamefont
  {Trivedi}}, \ and\ \bibinfo {author} {\bibfnamefont {A.}~\bibnamefont
  {Raghav}},\ }\href {\doibase 10.1103/PhysRevC.82.054601} {\bibfield
  {journal} {\bibinfo  {journal} {Physical Review C}\ }\textbf {\bibinfo
  {volume} {82}},\ \bibinfo {pages} {054601} (\bibinfo {year}
  {2010})}\BibitemShut {NoStop}%
\bibitem [{\citenamefont {Fang}\ \emph {et~al.}(2013)\citenamefont {Fang},
  \citenamefont {Gomes}, \citenamefont {Lubian}, \citenamefont {Zhou},
  \citenamefont {Zhang}, \citenamefont {Han}, \citenamefont {Liu},
  \citenamefont {Zheng}, \citenamefont {Guo}, \citenamefont {Wang},
  \citenamefont {Qiang}, \citenamefont {Wang}, \citenamefont {Wu},
  \citenamefont {He}, , \citenamefont {Zheng}, \citenamefont {Li},
  \citenamefont {Hu},\ and\ \citenamefont {Yao}}]{Fang2013}%
  \BibitemOpen
  \bibfield  {author} {\bibinfo {author} {\bibfnamefont {Y.~D.}\ \bibnamefont
  {Fang}}, \bibinfo {author} {\bibfnamefont {P.~R.~S.}\ \bibnamefont {Gomes}},
  \bibinfo {author} {\bibfnamefont {J.}~\bibnamefont {Lubian}}, \bibinfo
  {author} {\bibfnamefont {X.~H.}\ \bibnamefont {Zhou}}, \bibinfo {author}
  {\bibfnamefont {Y.~H.}\ \bibnamefont {Zhang}}, \bibinfo {author}
  {\bibfnamefont {J.~L.}\ \bibnamefont {Han}}, \bibinfo {author} {\bibfnamefont
  {M.~L.}\ \bibnamefont {Liu}}, \bibinfo {author} {\bibfnamefont
  {Y.}~\bibnamefont {Zheng}}, \bibinfo {author} {\bibfnamefont
  {S.}~\bibnamefont {Guo}}, \bibinfo {author} {\bibfnamefont {J.~G.}\
  \bibnamefont {Wang}}, \bibinfo {author} {\bibfnamefont {Y.~H.}\ \bibnamefont
  {Qiang}}, \bibinfo {author} {\bibfnamefont {Z.~G.}\ \bibnamefont {Wang}},
  \bibinfo {author} {\bibfnamefont {X.~G.}\ \bibnamefont {Wu}}, \bibinfo
  {author} {\bibfnamefont {C.~Y.}\ \bibnamefont {He}}, , \bibinfo {author}
  {\bibfnamefont {Y.}~\bibnamefont {Zheng}}, \bibinfo {author} {\bibfnamefont
  {C.~B.}\ \bibnamefont {Li}}, \bibinfo {author} {\bibfnamefont {S.~P.}\
  \bibnamefont {Hu}}, \ and\ \bibinfo {author} {\bibfnamefont {S.~H.}\
  \bibnamefont {Yao}},\ }\href {\doibase 10.1103/PhysRevC.87.024604} {\bibfield
   {journal} {\bibinfo  {journal} {Physical Review C}\ }\textbf {\bibinfo
  {volume} {87}},\ \bibinfo {pages} {024604} (\bibinfo {year}
  {2013})}\BibitemShut {NoStop}%
\bibitem [{\citenamefont {Hinde}\ \emph {et~al.}(2002)\citenamefont {Hinde},
  \citenamefont {Dasgupta}, \citenamefont {Fulton}, \citenamefont {Morton},
  \citenamefont {Woolliscroft}, \citenamefont {Berriman},\ and\ \citenamefont
  {Hagino}}]{Hinde2002}%
  \BibitemOpen
  \bibfield  {author} {\bibinfo {author} {\bibfnamefont {D.~J.}\ \bibnamefont
  {Hinde}}, \bibinfo {author} {\bibfnamefont {M.}~\bibnamefont {Dasgupta}},
  \bibinfo {author} {\bibfnamefont {B.~R.}\ \bibnamefont {Fulton}}, \bibinfo
  {author} {\bibfnamefont {C.~R.}\ \bibnamefont {Morton}}, \bibinfo {author}
  {\bibfnamefont {R.~J.}\ \bibnamefont {Woolliscroft}}, \bibinfo {author}
  {\bibfnamefont {A.~C.}\ \bibnamefont {Berriman}}, \ and\ \bibinfo {author}
  {\bibfnamefont {K.}~\bibnamefont {Hagino}},\ }\href {\doibase
  10.1103/PhysRevLett.89.272701} {\bibfield  {journal} {\bibinfo  {journal}
  {Physical Review Letters}\ }\textbf {\bibinfo {volume} {89}},\ \bibinfo
  {pages} {272701} (\bibinfo {year} {2002})}\BibitemShut {NoStop}%
\bibitem [{\citenamefont {Rafiei}\ \emph {et~al.}(2010)\citenamefont {Rafiei},
  \citenamefont {du~Rietz}, \citenamefont {Luong}, \citenamefont {Hinde},
  \citenamefont {Dasgupta}, \citenamefont {Evers},\ and\ \citenamefont
  {Diaz-Torres}}]{Rafiei2010b}%
  \BibitemOpen
  \bibfield  {author} {\bibinfo {author} {\bibfnamefont {R.}~\bibnamefont
  {Rafiei}}, \bibinfo {author} {\bibfnamefont {R.}~\bibnamefont {du~Rietz}},
  \bibinfo {author} {\bibfnamefont {D.~H.}\ \bibnamefont {Luong}}, \bibinfo
  {author} {\bibfnamefont {D.~J.}\ \bibnamefont {Hinde}}, \bibinfo {author}
  {\bibfnamefont {M.}~\bibnamefont {Dasgupta}}, \bibinfo {author}
  {\bibfnamefont {M.}~\bibnamefont {Evers}}, \ and\ \bibinfo {author}
  {\bibfnamefont {A.}~\bibnamefont {Diaz-Torres}},\ }\href {\doibase
  10.1103/PhysRevC.81.024601} {\bibfield  {journal} {\bibinfo  {journal}
  {Physical Review C}\ }\textbf {\bibinfo {volume} {81}},\ \bibinfo {pages}
  {024601} (\bibinfo {year} {2010})}\BibitemShut {NoStop}%
\bibitem [{\citenamefont {Luong}\ \emph {et~al.}(2011)\citenamefont {Luong},
  \citenamefont {Dasgupta}, \citenamefont {Hinde}, \citenamefont {{Du Rietz}},
  \citenamefont {Rafiei}, \citenamefont {Lin}, \citenamefont {Evers},\ and\
  \citenamefont {Diaz-Torres}}]{Luong2011b}%
  \BibitemOpen
  \bibfield  {author} {\bibinfo {author} {\bibfnamefont {D.~H.}\ \bibnamefont
  {Luong}}, \bibinfo {author} {\bibfnamefont {M.}~\bibnamefont {Dasgupta}},
  \bibinfo {author} {\bibfnamefont {D.~J.}\ \bibnamefont {Hinde}}, \bibinfo
  {author} {\bibfnamefont {R.}~\bibnamefont {{Du Rietz}}}, \bibinfo {author}
  {\bibfnamefont {R.}~\bibnamefont {Rafiei}}, \bibinfo {author} {\bibfnamefont
  {C.~J.}\ \bibnamefont {Lin}}, \bibinfo {author} {\bibfnamefont
  {M.}~\bibnamefont {Evers}}, \ and\ \bibinfo {author} {\bibfnamefont
  {A.}~\bibnamefont {Diaz-Torres}},\ }\href {\doibase
  10.1016/j.physletb.2010.11.007} {\bibfield  {journal} {\bibinfo  {journal}
  {Physics Letters B}\ }\textbf {\bibinfo {volume} {695}},\ \bibinfo {pages}
  {105} (\bibinfo {year} {2011})}\BibitemShut {NoStop}%
\bibitem [{\citenamefont {Tilley}\ \emph {et~al.}(2004)\citenamefont {Tilley},
  \citenamefont {Kelley}, \citenamefont {Godwin}, \citenamefont {Millener},
  \citenamefont {Purcell}, \citenamefont {Sheu},\ and\ \citenamefont
  {Weller}}]{Tilley2004}%
  \BibitemOpen
  \bibfield  {author} {\bibinfo {author} {\bibfnamefont {D.~R.}\ \bibnamefont
  {Tilley}}, \bibinfo {author} {\bibfnamefont {J.~H.}\ \bibnamefont {Kelley}},
  \bibinfo {author} {\bibfnamefont {J.~L.}\ \bibnamefont {Godwin}}, \bibinfo
  {author} {\bibfnamefont {D.~J.}\ \bibnamefont {Millener}}, \bibinfo {author}
  {\bibfnamefont {J.~E.}\ \bibnamefont {Purcell}}, \bibinfo {author}
  {\bibfnamefont {C.~G.}\ \bibnamefont {Sheu}}, \ and\ \bibinfo {author}
  {\bibfnamefont {H.~R.}\ \bibnamefont {Weller}},\ }\href {\doibase
  10.1016/j.nuclphysa.2004.09.059} {\bibfield  {journal} {\bibinfo  {journal}
  {Nuclear Physics A}\ }\textbf {\bibinfo {volume} {745}},\ \bibinfo {pages}
  {155} (\bibinfo {year} {2004})}\BibitemShut {NoStop}%
\bibitem [{\citenamefont {Diaz-Torres}\ \emph {et~al.}(2007)\citenamefont
  {Diaz-Torres}, \citenamefont {Hinde}, \citenamefont {Tostevin}, \citenamefont
  {Dasgupta},\ and\ \citenamefont {Gasques}}]{Diaz-Torres2007}%
  \BibitemOpen
  \bibfield  {author} {\bibinfo {author} {\bibfnamefont {A.}~\bibnamefont
  {Diaz-Torres}}, \bibinfo {author} {\bibfnamefont {D.~J.}\ \bibnamefont
  {Hinde}}, \bibinfo {author} {\bibfnamefont {J.~A.}\ \bibnamefont {Tostevin}},
  \bibinfo {author} {\bibfnamefont {M.}~\bibnamefont {Dasgupta}}, \ and\
  \bibinfo {author} {\bibfnamefont {L.~R.}\ \bibnamefont {Gasques}},\ }\href
  {\doibase 10.1103/PhysRevLett.98.152701} {\bibfield  {journal} {\bibinfo
  {journal} {Physical Review Letters}\ }\textbf {\bibinfo {volume} {98}},\
  \bibinfo {pages} {152701} (\bibinfo {year} {2007})}\BibitemShut {NoStop}%
\bibitem [{\citenamefont {Diaz-Torres}(2011)}]{Diaz-Torres2011}%
  \BibitemOpen
  \bibfield  {author} {\bibinfo {author} {\bibfnamefont {A.}~\bibnamefont
  {Diaz-Torres}},\ }\href {\doibase 10.1016/j.cpc.2010.12.053} {\bibfield
  {journal} {\bibinfo  {journal} {Computer Physics Communications}\ }\textbf
  {\bibinfo {volume} {182}},\ \bibinfo {pages} {1100} (\bibinfo {year}
  {2011})}\BibitemShut {NoStop}%
\bibitem [{\citenamefont {Simpson}\ \emph {et~al.}(2016)\citenamefont
  {Simpson}, \citenamefont {Cook}, \citenamefont {Luong}, \citenamefont
  {Kalkal}, \citenamefont {Carter}, \citenamefont {Dasgupta}, \citenamefont
  {Hinde},\ and\ \citenamefont {Williams}}]{simpson2016}%
  \BibitemOpen
  \bibfield  {author} {\bibinfo {author} {\bibfnamefont {E.~C.}\ \bibnamefont
  {Simpson}}, \bibinfo {author} {\bibfnamefont {K.~J.}\ \bibnamefont {Cook}},
  \bibinfo {author} {\bibfnamefont {D.~H.}\ \bibnamefont {Luong}}, \bibinfo
  {author} {\bibfnamefont {S.}~\bibnamefont {Kalkal}}, \bibinfo {author}
  {\bibfnamefont {I.~P.}\ \bibnamefont {Carter}}, \bibinfo {author}
  {\bibfnamefont {M.}~\bibnamefont {Dasgupta}}, \bibinfo {author}
  {\bibfnamefont {D.~J.}\ \bibnamefont {Hinde}}, \ and\ \bibinfo {author}
  {\bibfnamefont {E.}~\bibnamefont {Williams}},\ }\href {\doibase
  10.1103/PhysRevC.93.024605} {\bibfield  {journal} {\bibinfo  {journal}
  {Physical Review C}\ }\textbf {\bibinfo {volume} {93}},\ \bibinfo {pages}
  {024605} (\bibinfo {year} {2016})}\BibitemShut {NoStop}%
\bibitem [{\citenamefont {Barker}\ and\ \citenamefont
  {Treacy}(1962)}]{Barker1962}%
  \BibitemOpen
  \bibfield  {author} {\bibinfo {author} {\bibfnamefont {F.~C.}\ \bibnamefont
  {Barker}}\ and\ \bibinfo {author} {\bibfnamefont {P.~B.}\ \bibnamefont
  {Treacy}},\ }\href@noop {} {\bibfield  {journal} {\bibinfo  {journal}
  {Nuclear Physics}\ }\textbf {\bibinfo {volume} {38}},\ \bibinfo {pages} {33}
  (\bibinfo {year} {1962})}\BibitemShut {NoStop}%
\bibitem [{\citenamefont {Barker}(1988)}]{Barker1988}%
  \BibitemOpen
  \bibfield  {author} {\bibinfo {author} {\bibfnamefont {F.~C.}\ \bibnamefont
  {Barker}},\ }\href@noop {} {\bibfield  {journal} {\bibinfo  {journal}
  {Australian Journal of Physics}\ }\textbf {\bibinfo {volume} {41}},\ \bibinfo
  {pages} {743} (\bibinfo {year} {1988})}\BibitemShut {NoStop}%
\bibitem [{\citenamefont {Luong}\ \emph {et~al.}(2013)\citenamefont {Luong},
  \citenamefont {Dasgupta}, \citenamefont {Hinde}, \citenamefont {du~Rietz},
  \citenamefont {Rafiei}, \citenamefont {Lin}, \citenamefont {Evers},\ and\
  \citenamefont {Diaz-Torres}}]{Luong2013}%
  \BibitemOpen
  \bibfield  {author} {\bibinfo {author} {\bibfnamefont {D.~H.}\ \bibnamefont
  {Luong}}, \bibinfo {author} {\bibfnamefont {M.}~\bibnamefont {Dasgupta}},
  \bibinfo {author} {\bibfnamefont {D.~J.}\ \bibnamefont {Hinde}}, \bibinfo
  {author} {\bibfnamefont {R.}~\bibnamefont {du~Rietz}}, \bibinfo {author}
  {\bibfnamefont {R.}~\bibnamefont {Rafiei}}, \bibinfo {author} {\bibfnamefont
  {C.~J.}\ \bibnamefont {Lin}}, \bibinfo {author} {\bibfnamefont
  {M.}~\bibnamefont {Evers}}, \ and\ \bibinfo {author} {\bibfnamefont
  {A.}~\bibnamefont {Diaz-Torres}},\ }\href {\doibase
  10.1103/PhysRevC.88.034609} {\bibfield  {journal} {\bibinfo  {journal} {Phys.
  Rev. C}\ }\textbf {\bibinfo {volume} {88}},\ \bibinfo {pages} {34609}
  (\bibinfo {year} {2013})}\BibitemShut {NoStop}%
\bibitem [{\citenamefont {McIntosh}\ \emph {et~al.}(2007)\citenamefont
  {McIntosh}, \citenamefont {Hudan}, \citenamefont {Metelko}, \citenamefont
  {{De Souza}}, \citenamefont {Charity}, \citenamefont {Sobotka}, \citenamefont
  {Lynch},\ and\ \citenamefont {Tsang}}]{McIntosh2007}%
  \BibitemOpen
  \bibfield  {author} {\bibinfo {author} {\bibfnamefont {A.~B.}\ \bibnamefont
  {McIntosh}}, \bibinfo {author} {\bibfnamefont {S.}~\bibnamefont {Hudan}},
  \bibinfo {author} {\bibfnamefont {C.~J.}\ \bibnamefont {Metelko}}, \bibinfo
  {author} {\bibfnamefont {R.~T.}\ \bibnamefont {{De Souza}}}, \bibinfo
  {author} {\bibfnamefont {R.~J.}\ \bibnamefont {Charity}}, \bibinfo {author}
  {\bibfnamefont {L.~G.}\ \bibnamefont {Sobotka}}, \bibinfo {author}
  {\bibfnamefont {W.~G.}\ \bibnamefont {Lynch}}, \ and\ \bibinfo {author}
  {\bibfnamefont {M.~B.}\ \bibnamefont {Tsang}},\ }\href {\doibase
  10.1103/PhysRevLett.99.132701} {\bibfield  {journal} {\bibinfo  {journal}
  {Physical Review Letters}\ }\textbf {\bibinfo {volume} {99}},\ \bibinfo
  {pages} {132701} (\bibinfo {year} {2007})}\BibitemShut {NoStop}%
\bibitem [{\citenamefont {Kalkal}\ \emph {et~al.}(2016)\citenamefont {Kalkal},
  \citenamefont {Simpson}, \citenamefont {Luong}, \citenamefont {Cook},
  \citenamefont {Dasgupta}, \citenamefont {Hinde}, \citenamefont {Carter},
  \citenamefont {Jeung}, \citenamefont {Mohanto}, \citenamefont {Palshetkar},
  \citenamefont {Prasad}, \citenamefont {Rafferty}, \citenamefont {Simenel},
  \citenamefont {Vo-Phuoc}, \citenamefont {Williams}, \citenamefont {Gasques},
  \citenamefont {Gomes},\ and\ \citenamefont {Linares}}]{kalkal2016}%
  \BibitemOpen
  \bibfield  {author} {\bibinfo {author} {\bibfnamefont {S.}~\bibnamefont
  {Kalkal}}, \bibinfo {author} {\bibfnamefont {E.~C.}\ \bibnamefont {Simpson}},
  \bibinfo {author} {\bibfnamefont {D.~H.}\ \bibnamefont {Luong}}, \bibinfo
  {author} {\bibfnamefont {K.~J.}\ \bibnamefont {Cook}}, \bibinfo {author}
  {\bibfnamefont {M.}~\bibnamefont {Dasgupta}}, \bibinfo {author}
  {\bibfnamefont {D.~J.}\ \bibnamefont {Hinde}}, \bibinfo {author}
  {\bibfnamefont {I.~P.}\ \bibnamefont {Carter}}, \bibinfo {author}
  {\bibfnamefont {D.~Y.}\ \bibnamefont {Jeung}}, \bibinfo {author}
  {\bibfnamefont {G.}~\bibnamefont {Mohanto}}, \bibinfo {author} {\bibfnamefont
  {C.~S.}\ \bibnamefont {Palshetkar}}, \bibinfo {author} {\bibfnamefont
  {E.}~\bibnamefont {Prasad}}, \bibinfo {author} {\bibfnamefont {D.~C.}\
  \bibnamefont {Rafferty}}, \bibinfo {author} {\bibfnamefont {C.}~\bibnamefont
  {Simenel}}, \bibinfo {author} {\bibfnamefont {K.}~\bibnamefont {Vo-Phuoc}},
  \bibinfo {author} {\bibfnamefont {E.}~\bibnamefont {Williams}}, \bibinfo
  {author} {\bibfnamefont {L.~R.}\ \bibnamefont {Gasques}}, \bibinfo {author}
  {\bibfnamefont {P.~R.~S.}\ \bibnamefont {Gomes}}, \ and\ \bibinfo {author}
  {\bibfnamefont {R.}~\bibnamefont {Linares}},\ }\href {\doibase
  10.1103/PhysRevC.93.044605} {\bibfield  {journal} {\bibinfo  {journal} {Phys.
  Rev. C}\ }\textbf {\bibinfo {volume} {93}},\ \bibinfo {pages} {044605}
  (\bibinfo {year} {2016})}\BibitemShut {NoStop}%
\bibitem [{\citenamefont {Mason}\ \emph {et~al.}(1992)\citenamefont {Mason},
  \citenamefont {Gazes}, \citenamefont {Roberts},\ and\ \citenamefont
  {Teichmann}}]{mason1992}%
  \BibitemOpen
  \bibfield  {author} {\bibinfo {author} {\bibfnamefont {J.~E.}\ \bibnamefont
  {Mason}}, \bibinfo {author} {\bibfnamefont {S.~B.}\ \bibnamefont {Gazes}},
  \bibinfo {author} {\bibfnamefont {R.~B.}\ \bibnamefont {Roberts}}, \ and\
  \bibinfo {author} {\bibfnamefont {S.~G.}\ \bibnamefont {Teichmann}},\
  }\href@noop {} {\bibfield  {journal} {\bibinfo  {journal} {Physical Review
  C}\ }\textbf {\bibinfo {volume} {45}},\ \bibinfo {pages} {2870} (\bibinfo
  {year} {1992})}\BibitemShut {NoStop}%
\bibitem [{\citenamefont {Corradi}\ \emph {et~al.}(1990)\citenamefont
  {Corradi}, \citenamefont {Skorka}, \citenamefont {Lenz}, \citenamefont
  {L\"{o}bner}, \citenamefont {Pascholati}, \citenamefont {Quade},
  \citenamefont {Rudolph}, \citenamefont {Schomburg}, \citenamefont
  {Steinmayer}, \citenamefont {Theis}, \citenamefont {Montagnoli},
  \citenamefont {Napoli}, \citenamefont {Stefanini}, \citenamefont {Tivelli},
  \citenamefont {Beghini}, \citenamefont {Scarlassara}, \citenamefont
  {Signorini},\ and\ \citenamefont {Soramel}}]{Corradi1990}%
  \BibitemOpen
  \bibfield  {author} {\bibinfo {author} {\bibfnamefont {L.}~\bibnamefont
  {Corradi}}, \bibinfo {author} {\bibfnamefont {S.~J.}\ \bibnamefont {Skorka}},
  \bibinfo {author} {\bibfnamefont {U.}~\bibnamefont {Lenz}}, \bibinfo {author}
  {\bibfnamefont {K.~E.~G.}\ \bibnamefont {L\"{o}bner}}, \bibinfo {author}
  {\bibfnamefont {P.~R.}\ \bibnamefont {Pascholati}}, \bibinfo {author}
  {\bibfnamefont {U.}~\bibnamefont {Quade}}, \bibinfo {author} {\bibfnamefont
  {K.}~\bibnamefont {Rudolph}}, \bibinfo {author} {\bibfnamefont
  {W.}~\bibnamefont {Schomburg}}, \bibinfo {author} {\bibfnamefont
  {M.}~\bibnamefont {Steinmayer}}, \bibinfo {author} {\bibfnamefont {H.~G.}\
  \bibnamefont {Theis}}, \bibinfo {author} {\bibfnamefont {G.}~\bibnamefont
  {Montagnoli}}, \bibinfo {author} {\bibfnamefont {D.~R.}\ \bibnamefont
  {Napoli}}, \bibinfo {author} {\bibfnamefont {A.~M.}\ \bibnamefont
  {Stefanini}}, \bibinfo {author} {\bibfnamefont {A.}~\bibnamefont {Tivelli}},
  \bibinfo {author} {\bibfnamefont {S.}~\bibnamefont {Beghini}}, \bibinfo
  {author} {\bibfnamefont {F.}~\bibnamefont {Scarlassara}}, \bibinfo {author}
  {\bibfnamefont {C.}~\bibnamefont {Signorini}}, \ and\ \bibinfo {author}
  {\bibfnamefont {F.}~\bibnamefont {Soramel}},\ }\href {\doibase
  10.1007/BF01289348} {\bibfield  {journal} {\bibinfo  {journal} {Z. Phys. A}\
  }\textbf {\bibinfo {volume} {335}},\ \bibinfo {pages} {55} (\bibinfo {year}
  {1990})}\BibitemShut {NoStop}%
\bibitem [{\citenamefont {Knoll}\ and\ \citenamefont
  {Schaeffer}(1977)}]{Knoll1977}%
  \BibitemOpen
  \bibfield  {author} {\bibinfo {author} {\bibfnamefont {J.}~\bibnamefont
  {Knoll}}\ and\ \bibinfo {author} {\bibfnamefont {R.}~\bibnamefont
  {Schaeffer}},\ }\href {\doibase
  http://dx.doi.org/10.1016/0370-1573(77)90030-8} {\bibfield  {journal}
  {\bibinfo  {journal} {Physics Reports}\ }\textbf {\bibinfo {volume} {31}},\
  \bibinfo {pages} {159 } (\bibinfo {year} {1977})}\BibitemShut {NoStop}%
\bibitem [{\citenamefont {Vigezzi}\ and\ \citenamefont
  {Winther}(1989)}]{Vigezzi1989}%
  \BibitemOpen
  \bibfield  {author} {\bibinfo {author} {\bibfnamefont {E.}~\bibnamefont
  {Vigezzi}}\ and\ \bibinfo {author} {\bibfnamefont {A.}~\bibnamefont
  {Winther}},\ }\href {\doibase http://dx.doi.org/10.1016/0003-4916(89)90145-0}
  {\bibfield  {journal} {\bibinfo  {journal} {Annals of Physics}\ }\textbf
  {\bibinfo {volume} {192}},\ \bibinfo {pages} {432 } (\bibinfo {year}
  {1989})}\BibitemShut {NoStop}%
\bibitem [{\citenamefont {Gomes}\ \emph {et~al.}(2011)\citenamefont {Gomes},
  \citenamefont {Linares}, \citenamefont {Lubian}, \citenamefont {Lopes},
  \citenamefont {Cardozo}, \citenamefont {Pereira},\ and\ \citenamefont
  {Padron}}]{Gomes2011}%
  \BibitemOpen
  \bibfield  {author} {\bibinfo {author} {\bibfnamefont {P.~R.~S.}\
  \bibnamefont {Gomes}}, \bibinfo {author} {\bibfnamefont {R.}~\bibnamefont
  {Linares}}, \bibinfo {author} {\bibfnamefont {J.}~\bibnamefont {Lubian}},
  \bibinfo {author} {\bibfnamefont {C.~C.}\ \bibnamefont {Lopes}}, \bibinfo
  {author} {\bibfnamefont {E.~N.}\ \bibnamefont {Cardozo}}, \bibinfo {author}
  {\bibfnamefont {B.~H.~F.}\ \bibnamefont {Pereira}}, \ and\ \bibinfo {author}
  {\bibfnamefont {I.}~\bibnamefont {Padron}},\ }\href {\doibase
  10.1103/PhysRevC.84.014615} {\bibfield  {journal} {\bibinfo  {journal}
  {Physical Review C}\ }\textbf {\bibinfo {volume} {84}},\ \bibinfo {pages}
  {014615} (\bibinfo {year} {2011})}\BibitemShut {NoStop}%
\bibitem [{\citenamefont {Tripathi}\ \emph {et~al.}(2005)\citenamefont
  {Tripathi}, \citenamefont {Navin}, \citenamefont {Nanal}, \citenamefont
  {Pillay}, \citenamefont {Mahata}, \citenamefont {Ramachandran}, \citenamefont
  {Shrivastava}, \citenamefont {Chatterjee},\ and\ \citenamefont
  {Kailas}}]{Tripathi2005}%
  \BibitemOpen
  \bibfield  {author} {\bibinfo {author} {\bibfnamefont {V.}~\bibnamefont
  {Tripathi}}, \bibinfo {author} {\bibfnamefont {A.}~\bibnamefont {Navin}},
  \bibinfo {author} {\bibfnamefont {V.}~\bibnamefont {Nanal}}, \bibinfo
  {author} {\bibfnamefont {R.~G.}\ \bibnamefont {Pillay}}, \bibinfo {author}
  {\bibfnamefont {K.}~\bibnamefont {Mahata}}, \bibinfo {author} {\bibfnamefont
  {K.}~\bibnamefont {Ramachandran}}, \bibinfo {author} {\bibfnamefont
  {A.}~\bibnamefont {Shrivastava}}, \bibinfo {author} {\bibfnamefont
  {A.}~\bibnamefont {Chatterjee}}, \ and\ \bibinfo {author} {\bibfnamefont
  {S.}~\bibnamefont {Kailas}},\ }\href {\doibase 10.1103/PhysRevC.72.017601}
  {\bibfield  {journal} {\bibinfo  {journal} {Physical Review C}\ }\textbf
  {\bibinfo {volume} {72}},\ \bibinfo {pages} {1} (\bibinfo {year}
  {2005})}\BibitemShut {NoStop}%
\bibitem [{\citenamefont {Chamon}\ \emph {et~al.}(2002)\citenamefont {Chamon},
  \citenamefont {Carlson}, \citenamefont {Gasques}, \citenamefont {Pereira},
  \citenamefont {{De Conti}}, \citenamefont {Alvarez}, \citenamefont {Hussein},
  \citenamefont {{C{\^{a}}ndido Ribeiro}}, \citenamefont {Rossi},\ and\
  \citenamefont {Silva}}]{Chamon2002}%
  \BibitemOpen
  \bibfield  {author} {\bibinfo {author} {\bibfnamefont {L.~C.}\ \bibnamefont
  {Chamon}}, \bibinfo {author} {\bibfnamefont {B.~V.}\ \bibnamefont {Carlson}},
  \bibinfo {author} {\bibfnamefont {L.~R.}\ \bibnamefont {Gasques}}, \bibinfo
  {author} {\bibfnamefont {D.}~\bibnamefont {Pereira}}, \bibinfo {author}
  {\bibfnamefont {C.}~\bibnamefont {{De Conti}}}, \bibinfo {author}
  {\bibfnamefont {M.~A.~G.}\ \bibnamefont {Alvarez}}, \bibinfo {author}
  {\bibfnamefont {M.~S.}\ \bibnamefont {Hussein}}, \bibinfo {author}
  {\bibfnamefont {M.~A.}\ \bibnamefont {{C{\^{a}}ndido Ribeiro}}}, \bibinfo
  {author} {\bibfnamefont {E.~S.}\ \bibnamefont {Rossi}}, \ and\ \bibinfo
  {author} {\bibfnamefont {C.~P.}\ \bibnamefont {Silva}},\ }\href {\doibase
  10.1103/PhysRevC.66.014610} {\bibfield  {journal} {\bibinfo  {journal}
  {Physical Review C}\ }\textbf {\bibinfo {volume} {66}},\ \bibinfo {pages}
  {014610} (\bibinfo {year} {2002})}\BibitemShut {NoStop}%
\bibitem [{\citenamefont {Signorini}\ \emph {et~al.}(2002)\citenamefont
  {Signorini}, \citenamefont {Andrighetto}, \citenamefont {Guo}, \citenamefont
  {Ruan}, \citenamefont {Stroe}, \citenamefont {Soramel}, \citenamefont
  {L{\"{o}}bner}, \citenamefont {M{\"{u}}ller}, \citenamefont {Pierroutsakou},
  \citenamefont {Romoli}, \citenamefont {Rudolph}, \citenamefont {Thompson},
  \citenamefont {Trotta},\ and\ \citenamefont {Vitturi}}]{Signorini2002}%
  \BibitemOpen
  \bibfield  {author} {\bibinfo {author} {\bibfnamefont {C.}~\bibnamefont
  {Signorini}}, \bibinfo {author} {\bibfnamefont {A.}~\bibnamefont
  {Andrighetto}}, \bibinfo {author} {\bibfnamefont {J.~Y.}\ \bibnamefont
  {Guo}}, \bibinfo {author} {\bibfnamefont {M.}~\bibnamefont {Ruan}}, \bibinfo
  {author} {\bibfnamefont {L.}~\bibnamefont {Stroe}}, \bibinfo {author}
  {\bibfnamefont {F.}~\bibnamefont {Soramel}}, \bibinfo {author} {\bibfnamefont
  {K.~E.~G.}\ \bibnamefont {L{\"{o}}bner}}, \bibinfo {author} {\bibfnamefont
  {L.}~\bibnamefont {M{\"{u}}ller}}, \bibinfo {author} {\bibfnamefont
  {D.}~\bibnamefont {Pierroutsakou}}, \bibinfo {author} {\bibfnamefont
  {M.}~\bibnamefont {Romoli}}, \bibinfo {author} {\bibfnamefont
  {K.}~\bibnamefont {Rudolph}}, \bibinfo {author} {\bibfnamefont
  {I.}~\bibnamefont {Thompson}}, \bibinfo {author} {\bibfnamefont
  {M.}~\bibnamefont {Trotta}}, \ and\ \bibinfo {author} {\bibfnamefont
  {A.}~\bibnamefont {Vitturi}},\ }\href {\doibase
  10.1016/S0375-9474(01)01541-X} {\bibfield  {journal} {\bibinfo  {journal}
  {Nuclear Physics A}\ }\textbf {\bibinfo {volume} {701}},\ \bibinfo {pages}
  {23c} (\bibinfo {year} {2002})}\BibitemShut {NoStop}%
\bibitem [{\citenamefont {Parkar}\ \emph {et~al.}(2013)\citenamefont {Parkar},
  \citenamefont {Jha}, \citenamefont {Pandit}, \citenamefont {Santra},\ and\
  \citenamefont {Kailas}}]{Parkar2013}%
  \BibitemOpen
  \bibfield  {author} {\bibinfo {author} {\bibfnamefont {V.~V.}\ \bibnamefont
  {Parkar}}, \bibinfo {author} {\bibfnamefont {V.}~\bibnamefont {Jha}},
  \bibinfo {author} {\bibfnamefont {S.~K.}\ \bibnamefont {Pandit}}, \bibinfo
  {author} {\bibfnamefont {S.}~\bibnamefont {Santra}}, \ and\ \bibinfo {author}
  {\bibfnamefont {S.}~\bibnamefont {Kailas}},\ }\href {\doibase
  10.1103/PhysRevC.87.034602} {\bibfield  {journal} {\bibinfo  {journal}
  {Physical Review C}\ }\textbf {\bibinfo {volume} {87}},\ \bibinfo {pages}
  {034602} (\bibinfo {year} {2013})}\BibitemShut {NoStop}%
\bibitem [{\citenamefont {Yu}\ \emph {et~al.}(2010)\citenamefont {Yu},
  \citenamefont {Zhang}, \citenamefont {Jia}, \citenamefont {Zhang},
  \citenamefont {Ruan}, \citenamefont {Yang}, \citenamefont {Wu}, \citenamefont
  {Xu},\ and\ \citenamefont {Bai}}]{Yu2010}%
  \BibitemOpen
  \bibfield  {author} {\bibinfo {author} {\bibfnamefont {N.}~\bibnamefont
  {Yu}}, \bibinfo {author} {\bibfnamefont {H.~Q.}\ \bibnamefont {Zhang}},
  \bibinfo {author} {\bibfnamefont {H.~M.}\ \bibnamefont {Jia}}, \bibinfo
  {author} {\bibfnamefont {S.~T.}\ \bibnamefont {Zhang}}, \bibinfo {author}
  {\bibfnamefont {M.}~\bibnamefont {Ruan}}, \bibinfo {author} {\bibfnamefont
  {F.}~\bibnamefont {Yang}}, \bibinfo {author} {\bibfnamefont {Z.~D.}\
  \bibnamefont {Wu}}, \bibinfo {author} {\bibfnamefont {X.~X.}\ \bibnamefont
  {Xu}}, \ and\ \bibinfo {author} {\bibfnamefont {C.~L.}\ \bibnamefont {Bai}},\
  }\href {\doibase 10.1088/0954-3899/37/7/075108} {\bibfield  {journal}
  {\bibinfo  {journal} {Journal of Physics G: Nuclear and Particle Physics}\
  }\textbf {\bibinfo {volume} {37}},\ \bibinfo {pages} {075108} (\bibinfo
  {year} {2010})}\BibitemShut {NoStop}%
\bibitem [{\citenamefont {Zagrebaev}\ \emph {et~al.}()\citenamefont
  {Zagrebaev}, \citenamefont {Denikin}, \citenamefont {Alekseev}, \citenamefont
  {Karpov}, \citenamefont {Samarin}, \citenamefont {Naumenko},\ and\
  \citenamefont {Kozhin}}]{Zagrebaev}%
  \BibitemOpen
  \bibfield  {author} {\bibinfo {author} {\bibfnamefont {V.~I.}\ \bibnamefont
  {Zagrebaev}}, \bibinfo {author} {\bibfnamefont {A.~S.}\ \bibnamefont
  {Denikin}}, \bibinfo {author} {\bibfnamefont {A.~P.}\ \bibnamefont
  {Alekseev}}, \bibinfo {author} {\bibfnamefont {A.~V.}\ \bibnamefont
  {Karpov}}, \bibinfo {author} {\bibfnamefont {V.~V.}\ \bibnamefont {Samarin}},
  \bibinfo {author} {\bibfnamefont {M.~A.}\ \bibnamefont {Naumenko}}, \ and\
  \bibinfo {author} {\bibfnamefont {A.~Y.}\ \bibnamefont {Kozhin}},\ }\href
  {http://nrv.jinr.ru/nrv/} {\enquote {\bibinfo {title} {{OM Code, Nuclear
  Reactions Video Project [URL: http://nrv.jinr.ru/nrv/]}},}\ }\BibitemShut
  {NoStop}%
\bibitem [{\citenamefont {Schiffer}\ \emph {et~al.}(1973)\citenamefont
  {Schiffer}, \citenamefont {Kijrner}, \citenamefont {Siemssen}, \citenamefont
  {Jones},\ and\ \citenamefont {Schwarzschild}}]{Schiffer1973}%
  \BibitemOpen
  \bibfield  {author} {\bibinfo {author} {\bibfnamefont {J.~P.}\ \bibnamefont
  {Schiffer}}, \bibinfo {author} {\bibfnamefont {H.~J.}\ \bibnamefont
  {Kijrner}}, \bibinfo {author} {\bibfnamefont {R.~H.}\ \bibnamefont
  {Siemssen}}, \bibinfo {author} {\bibfnamefont {K.~W.}\ \bibnamefont {Jones}},
  \ and\ \bibinfo {author} {\bibfnamefont {A.}~\bibnamefont {Schwarzschild}},\
  }\href@noop {} {\bibfield  {journal} {\bibinfo  {journal} {Physics Letters
  B.}\ }\textbf {\bibinfo {volume} {44}},\ \bibinfo {pages} {47} (\bibinfo
  {year} {1973})}\BibitemShut {NoStop}%
\bibitem [{\citenamefont {Diaz-Torres}(2010)}]{Diaz-Torrespc}%
  \BibitemOpen
  \bibfield  {author} {\bibinfo {author} {\bibfnamefont {A.}~\bibnamefont
  {Diaz-Torres}},\ }\href@noop {} {\enquote {\bibinfo {title} {(private
  communication)},}\ } (\bibinfo {year} {2010})\BibitemShut {NoStop}%
\end{thebibliography}%

\end{document}